\documentclass{aa}

\usepackage{graphicx}
\DeclareGraphicsExtensions{.eps,.ps,.pdf,.png,.jpg,.gif}
\usepackage{natbib}
\usepackage{txfonts}

\newcommand{\pun}[1]{\mbox{\rm\,#1}} % Command used to write physical units

\newcommand{\logg}{\ensuremath{\log g}}

\newcommand{\mlp}{\ensuremath{\alpha_{\mathrm{MLT}}}}

\newcommand{\moh}{\ensuremath{[\mathrm{M/H}]}}
\newcommand{\feoh}{\ensuremath{[\mathrm{Fe/H}]}}

\newcommand{\Teff}{\ensuremath{T_{\mathrm{eff}}}}

\newcommand{\beq}{\begin{equation}}
\newcommand{\eeq}{\end{equation}}

\newcommand{\xtmean}[1]{\ensuremath{\left\langle #1\right\rangle}}

\newcommand{\eref}[1]{\mbox{(\ref{#1})}}

\newcommand{\cp}{\ensuremath{c_{\mathrm{p}}}}
\newcommand{\taueddy}{\ensuremath{\tau_{\mathrm{e}}}}

\newcommand{\lmix}{\ensuremath{\Lambda}}
\newcommand{\Hp}{\ensuremath{H_{\mathrm{P}}}}

\newcommand{\MoH}{\ensuremath{\left[\mathrm{M}/\mathrm{H}\right]}}

\newcommand{\COBOLD}{{\tt CO$^5$BOLD}}
\newcommand{\LHD}{{\tt LHD}}

\newcommand{\MARCS}{{\tt MARCS}}

\newcommand{\vover}{\ensuremath{v_\mathrm{over}}}

\newcommand{\Blam}{\ensuremath{B_\lambda}}
\newcommand{\klam}{\ensuremath{\kappa_\lambda}}
\newcommand{\intlam}{\ensuremath{\int\!d\lambda\,}}
\newcommand{\fom}{\ensuremath{f_\Omega}}
\newcommand{\fb}{\ensuremath{f_B}}
\newcommand{\dT}{\ensuremath{\Delta T}}

\newcommand{\change}[1]{#1}

\begin{document}

\title{3D hydrodynamical \COBOLD\ model atmospheres of red giant stars}

\subtitle{I. Atmospheric structure of a giant located near the RGB tip}

\author{Hans-G\"{u}nter Ludwig\inst{1}
        \and
        Ar\={u}nas Ku\v{c}inskas\inst{2,3}
%        \fnmsep\thanks{We love offprint requests, preferably accompanied with solid check in hard currency}
       }

\offprints{H.-G. Ludwig}

\institute{
        Zentrum f\"ur Astronomie der Universit\"at Heidelberg,
        Landessternwarte, K\"{o}nigstuhl 12, D-69117 Heidelberg, Germany\\
        \email{hludwig@lsw.uni-heidelberg.de}
        \and
        Vilnius University Institute of Theoretical Physics and Astronomy, A. Go\v {s}tauto 12, Vilnius LT-01108, Lithuania
        \and
        Vilnius University Astronomical Observatory, M. K. \v{C}iurlionio 29, Vilnius LT-10222, Lithuania\\
        \email{arunaskc@itpa.lt}
%        \thanks{Mail server at ITPA only accepts incoming mail on Fridays}
        }

\date{Received: date; accepted: date}

\authorrunning{H.-G. Ludwig and A. Ku\v{c}inskas}
\titlerunning{3D hydrodynamical \COBOLD\ model atmospheres of red giant stars. I.}

% \abstract{}{}{}{}{}
% 5 {} token are mandatory

\abstract
{Red giant stars are important tracers of stellar populations in the Galaxy and
  beyond, thus accurate modeling of their structure and related observable
  properties is of great importance. Three-dimensional (3D)
  hydrodynamical stellar atmosphere models offer a new level of realism in
  the modeling of red giant atmospheres but still need to be established
  as standard tools.}
{We investigate the character and role of convection in the atmosphere of a
  prototypical red giant located close to the red giant branch (RGB) tip
  with atmospheric parameters, $T_{\rm eff}=3660$\,K, $\log g =
  1.0$, $\moh=0.0$.}
{Differential analysis of the atmospheric
 structures is performed using the 3D hydrodynamical and 1D classical
 atmosphere models calculated with the \COBOLD\ and \LHD\ codes,
 respectively. All models share identical atmospheric parameters, elemental 
 composition, opacities and equation-of-state.}
{We find that the atmosphere of this particular red giant consists of two
  rather distinct regions: the lower atmosphere dominated by convective motions
  and the upper atmosphere dominated by wave activity. Convective motions form a
  prominent granulation pattern with an intensity contrast ($\sim18\%$) which
  is larger than in the solar models ($\sim15\%$). The upper atmosphere is
  frequently traversed by fast shock waves, with vertical and horizontal
  velocities of up to Mach\,$\sim2.5$ and $\sim6.0$, respectively. The typical
  diameter of the granules amounts to $\sim5$\,Gm which translates into $\sim400$
  granules covering the whole stellar surface. The turbulent pressure in the
  giant model contributes up to $\sim35\%$ to the total (i.e., gas plus turbulent)
  pressure which shows that it cannot be neglected in
  stellar atmosphere and evolutionary modeling. However, there exists no
  combination of the mixing-length parameter, \mlp, and turbulent pressure,
  $P_{\rm turb}$, that would allow to satisfactorily reproduce the 3D
  temperature-pressure profile with 1D atmosphere models based on a standard
  formulation of mixing-length theory.}
   {}


   \maketitle

%------------------------------------------------------------------------------
\section{Introduction}
%------------------------------------------------------------------------------

Convection plays an important role in governing the interior structure and
evolution of red giants (i.e., stars on the red and asymptotic giant
branches, RGB/AGB). Besides of aiding the energy transport from the stellar
interior to the outer layers, convection mixes heavy elements from the nuclear
burning layers up into the stellar envelope and atmosphere. Since convective
mixing changes the local chemical composition it alters the stellar structure
because of changes in the opacities, thermodynamic properties, and nuclear
reaction rates of the stellar plasma. This affects the observable properties
of a star, and a proper understanding of convection is thus of fundamental
importance for building realistic models for stellar structure and evolution,
which, in turn, are fundamental building blocks of our understanding of
individual stars and stellar populations.

Convection in current one-dimensional (1D) hydrostatic stellar atmosphere
models is treated in a simplified way, typically, using the classical
mixing-length theory \citep[MLT,][]{BV58} or one of its more advanced variants
\citep{CM91,CGM96}. This approach has a number of drawbacks. For instance, the
efficiency of convective transport in the framework of MLT is scaled by the
apriori unknown mixing-length parameter, \mlp. It is commonly taken as a
fixed ratio of the mixing-length to the local pressure scale height, and is
usually calibrated using theoretical models of the Sun. Since the MLT is a
rather simplistic approach, \mlp\ needs not to be the same in main-sequence
stars, subgiants, giants and supergiants, as is normally assumed in the
calculation of stellar atmosphere models \citep{CK03,BH05,GEK08} or stellar
evolutionary tracks and isochrones \citep{DWKY04,VBD06,DCJ08,BGMN08,BNGM09}.

Despite significant efforts made during the last few decades to improve the
treatment of convection in stellar structure models a fundamental breakthrough
is still missing.  Seeking numerical solutions of the underlying
radiation-hydrodynamical equations is a promising way to make progress since
such models lay-out a clear path towards increasing realism in the description
of convection. In this class of models three-dimensional (3D) hydrodynamical
model atmospheres have already demonstrated their excellent capabilities of
reproducing the observed properties of surface convection in the Sun
\citep[e.g.,][]{SN98}, and associated spectral diagnostics
\citep[e.g.,][]{ANT00,CLS08,CLS10}, in many cases outperforming classical 1D
models despite lacking tunable parameters as in 1D. Similarly successfully,
such 3D models were applied to other types of stars, such as late-type dwarfs
\citep[][]{GHB09,RAK09,BBL10} and subgiants \citep[][]{CAN09}.

The progress in the 3D modeling of red giant atmospheres has been
considerably slower. To large extent this is related to the fact that
giant models are computationally more demanding, and the calculation of a grid
of giant atmosphere models is a sizable task. Among the early efforts, a 3D
model atmosphere of a red giant was discussed by \citet{KHL05}. The
authors have compared broad band photometric colors as predicted by the
classical 1D and 3D hydrodynamical model atmospheres and have shown that for
certain color indices the 3D--1D differences may reach $\sim$\,0.25\,mag. A
detailed analysis of the spectral line formation in the atmospheres of
somewhat warmer red giants ($\Teff\approx4700-5100$\,K, $\logg=2.2$,
$\moh=0.0$ to $-3.0$) was carried out by \citet{CAT07}. An interesting
finding of this work is that convective motions may produce significantly
cooler average temperatures in the outer atmospheric layers, an
effect which is increasingly pronounced at low metallicities
($\feoh<-2.0$). Similar effects have been seen in the 3D atmosphere models of
late-type dwarfs too, see, e.g., \citet{ANTS99}, \citet{BBL10},
\citet{GBL10}. Spectral lines of various atoms and molecules appear typically
stronger in a 3D than in a 1D model which leads to a disagreement between the
chemical abundances, $\epsilon(X)$, reaching $\Delta\log \epsilon(X) =
-0.5\dots\ -1.0$\,dex at the lowest metallicities.
Despite these successful attempts, 3D hydrodynamical model atmospheres of red
giants still need to be consolidated including comparisons to observations as
exemplified by \citet{RCL10} who performed a study of a metal-poor giant with
emphasis on hydrodynamic properties.

To further broaden the 3D model basis of giants, we undertook a study of the
influence of convection on the structure and observable properties of red
giants. The models were calculated with the 3D radiation-hydrodynamics code
\COBOLD\ and will eventually cover the entire range of stellar parameters
typical for stars on the red and asymptotic giant branches (RGB and AGB,
respectively). Some of these models are already available as part of the
CIFIST grid of \COBOLD\ 3D model atmospheres \citep{LCS09}. This homogeneous 
set of 3D model atmospheres is well suited to investigate the role of 
convection in the atmospheres of red giants of different effective 
temperatures, gravities and metallicities.

This paper summarizes the first results of the project, focusing on the role
of convection shaping the atmospheric structure of a solar-metallicity red
giant located close to the RGB tip.  The atmospheric parameters of the model
are \Teff=3660\,K, \logg=1.0, \MoH=0.0. The analysis is done differentially by
comparing 3D hydrodynamical and 1D static model atmospheres calculated for the
same set of atmospheric parameters and with identical opacities and
equation-of-state. In two companion papers we will further discuss the effects
of convection on the observable properties of this particular red giant, by
taking a closer look at the formation of individual spectral lines and global
properties of the spectral energy distribution (Ku\v{c}inskas et
al. 2012a,b\,, in prep.) The present paper mostly describes morphological
properties of the particular model which we, however, generalize in places to
make statements about convection in red giants in general.

%------------------------------------------------------------------------
\section{The model setup\label{s:model}}
%------------------------------------------------------------------------

\begin{figure*}[!t]
\centering
\resizebox{\hsize}{!}{\includegraphics{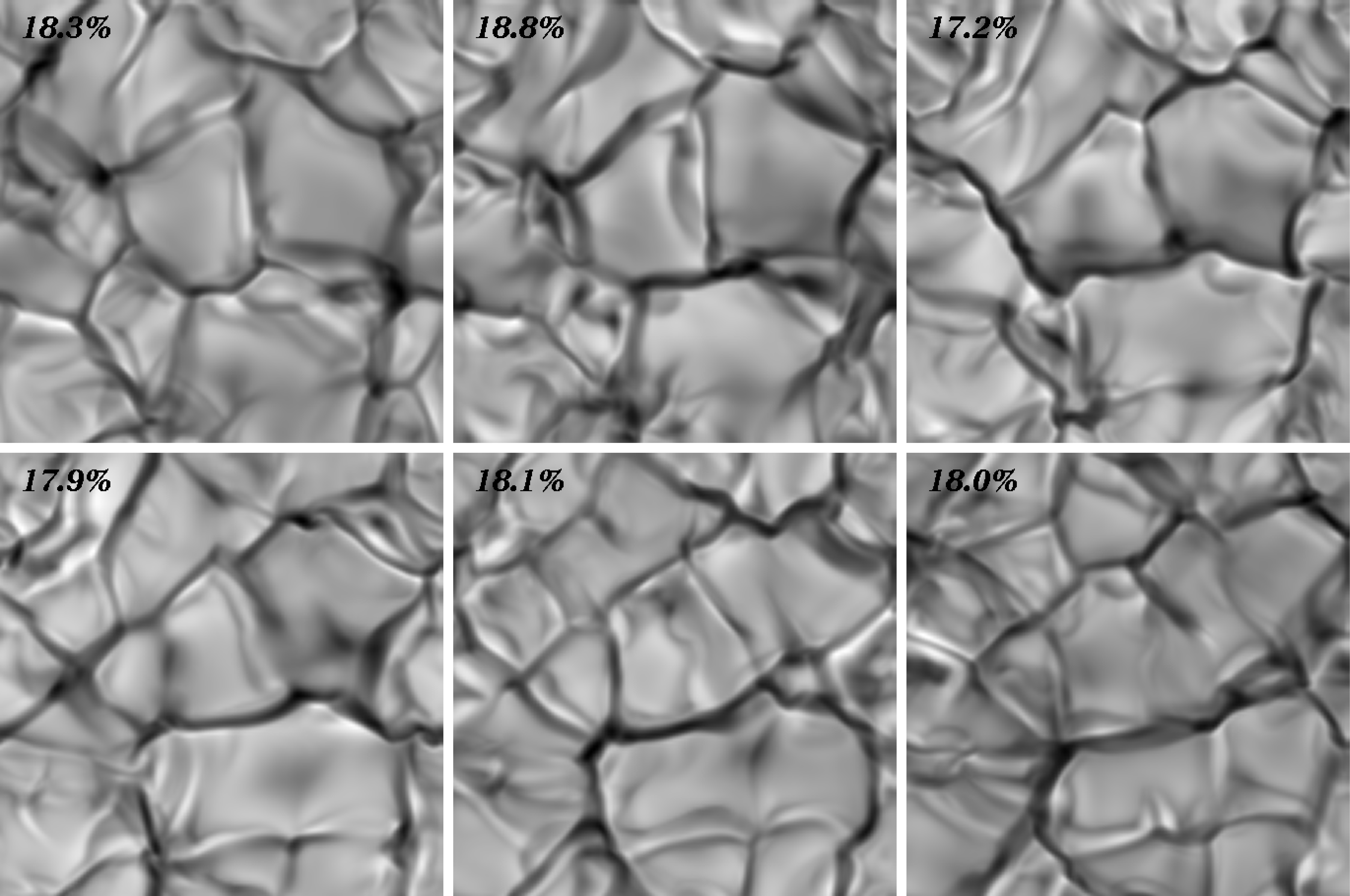}}
\caption{Snapshots of the emergent white light intensity during the temporal
  evolution (ordered upper-left to lower-right) of the hydrodynamical red
  giant model atmosphere. Bright granules where the matter is rising are
  surrounded by darker and significantly narrower intergranular lanes
  associated with down-flows. The spatial size of each frame is
  \mbox{15.6$\times$15.6\pun{Gm}$^2$}, and the time interval between frames
  300\pun{ks}. The relative RMS intensity contrast is given in the upper left
  corner of each frame.
  \label{f:timeseries}}
\end{figure*}

Two stellar atmosphere models were used in this study to assess the influence
of convection on the atmospheric structures of a red giant: 3D hydrodynamical
and classical 1D models, calculated with the codes \COBOLD\ and \LHD,
respectively. Both models were computed using identical atmospheric
parameters, $\Teff\approx3660$\,K, $\logg=1.0$, $\moh=0.0$. According to 
theoretical evolutionary tracks \citep[e.g.,][]{CPC07} this set of 
atmospheric parameters characterizes a late-type giant located close to the 
RGB tip, with a mass of $\sim$\,2\,${\rm M_{\odot}}$ and age of 
{\mbox{$\sim$\,1\,Gyr}}. In turn, the mass and gravity give a radius of 
{\mbox{$\sim\,75\,{\rm R_{\odot}}$}}.

\subsection{The 3D \COBOLD\ model\label{s:cobold_model}}

The 3D atmosphere model was calculated using the
\COBOLD\ stellar atmosphere code. The \COBOLD\ code employs a Riemann solver
of Roe type to integrate the equations of hydrodynamics and simultaneously solve
for the frequency-dependent radiation field on a 3D Cartesian grid
\citep{FSD02,FSWL03,WFSLH04,FAL10}. For a detailed description of the code and
its applications see \citet{FSL12}.

The code was employed in the 'box-in-a-star' set-up using a grid of
$150\times150\times151$ mesh points ($x\times y\times z$, where $z$ is
vertical dimension), with a corresponding size of the computational domain of
$15.6\times15.6\times8.6$\,Gm$^3$. Effects related to the stellar sphericity
are neglected. We applied open boundary conditions in the vertical direction
and periodic boundaries in the horizontal direction. The convective flux at
the lower boundary was controlled by specifying the entropy of the inflowing
gas. The model spans a range in optical depth of
$-6.5\lesssim\log\,\langle\tau_{\rm Ross}\rangle \lesssim5$ and samples the
atmospheric layers over $\sim$11 pressure scale heights.

In the model calculations we used monochromatic opacities from the
\MARCS\ stellar atmosphere package \citep{GEK08}, grouped into 5 opacity bins
\citep[for more on opacity grouping scheme
  see][]{N82,L92,LJS94,VBS04}. Opacities were calculated assuming solar
elemental abundances according to \citet{GS98}, with the exception of carbon,
nitrogen and oxygen, for which the following values were used: A(C)=8.41,
A(N)=7.8, A(O)=8.67, similar to the CNO abundances
recommended by \citet{AGS05}. Local thermodynamical equilibrium (LTE) was
assumed throughout the entire atmosphere and scattering was treated as true
absorption.

The equation-of-state (EOS) used in the model simulations takes into account
the ionization of hydrogen and helium, as well as formation of H$_2$ molecules
according to Saha-Boltzmann statistics. The ionization of metals is ignored in
the current version of EOS since its importance for the
gross thermodynamical properties is minor.

After initial relaxation to a quasi-stationary state, the model simulations
were run to cover a span of $\sim6\times10^6$\,sec ($\sim 70$\,days) in
stellar time. This corresponds to $\sim7$ convective turnover times as
measured by the Brunt-V\"{a}is\"{a}l\"{a} and/or advection timescales, the latter
equal to the time needed by the convective material to cross one
mixing-length, $\Lambda\equiv\alpha_{\rm MLT}H_{\rm p}$, where $H_{\rm p}$ is
the pressure scale height (both timescales were estimated at $\tau_{\rm
  Ross}=1.0$, see Sect.~\ref{s:timescales}). A set of 70 relaxed 3D model
snapshots was used in our analysis of the atmospheric structure. \change{This
  is sufficient for the aspects addressed in this work. However, one should
  keep in mind that the
  overall statistics gathered is limited  (in part
  dictated by computational cost), and the precision of convection-related
  properties (e.g., the turbulent pressure) is likely not better than $\sim5\,\%$.}

\subsection{The average \xtmean{\mbox{3D}} \COBOLD\ model}

Since the 3D hydrodynamical and 1D classical model atmospheres are based on
different physical assumptions, one may naturally expect differences in their
resulting properties. Two aspects are worth mentioning
here. First, the 3D hydrodynamical model atmospheres predict significant
horizontal fluctuations in temperature, density, pressure, velocity and other
kinematic and thermodynamic properties. This may produce horizontal
variations in, e.g., shapes and strengths of spectral lines formed in
different parts of the model photosphere. Second, convection is a
natural process arising in hydrodynamical model atmospheres inherent to the
equations of hydrodynamics and radiative transfer. Besides convection as such,
hydrodynamical models exhibit significant overshoot of material into the upper
layers of the atmosphere which should be convectively stable under the
Schwarzschild criterium. None of these effects is properly accounted
for in the 1D classical models built on the prescription given by
MLT. The consequence is that the structure of 3D hydrodynamical
models, even if horizontally averaged, is different from
1D models calculated with identical atmospheric parameters.

The relative importance of these effects can be assessed using the average 3D
model. Such a model can be calculated by horizontally averaging the thermal
structure over several instances in time (snapshots). By definition, the new
one-dimensional structure, further referred to as \xtmean{\mbox{3D}} model,
retains no information about the horizontal inhomogeneities but keeps the
imprints from the different (i.e., hydrodynamical) treatment of convection in
the original 3D model. The 3D--\xtmean{\mbox{3D}} and \xtmean{\mbox{3D}}--1D
differences may therefore provide valuable information about the relative
importance of horizontal fluctuations and the time-dependent effects related
with the treatment of convection, respectively.

The \xtmean{\mbox{3D}} model used in this work was obtained by averaging 70 3D
snapshots over the surfaces of equal Rosseland optical depth. To preserve the
radiative properties of the original 3D model we averaged the fourth moment of
temperature and first moment of gas pressure, following the prescription given
in \citet{SLF95}.

\subsection{1D \LHD\ model}

Comparison 1D hydrostatic models were calculated with the \LHD\ code, using
the same atmospheric parameters, elemental abundances, opacities and EOS as in
the 3D model calculations described above. Convection in the \LHD\ models is
described using the mixing-length theory, adopting the formulation of
\citet{M78}. The \LHD\ models characterized by several different mixing-length
parameters were used in this work, with $\mlp=1.0, 1.5$ and 2.0. It should be
stressed though that the choice of the mixing-length parameter affects mostly
the deeper atmospheric layers and has limited influence on the formation
of the emitted spectrum in this particular red giant \citep[][Ku\v{c}inskas et
  al. 2012, submitted to A\&A]{KHL05}.

In a few cases we added an ad-hoc overshooting in the LHD models. This was
done by preventing the convective velocity going to zero in the formally
stable regions but forcing it to a prescribed, fixed value. The convective
flux was calculated from the standard MLT formulae. In the overshoot region
this results in a downward directed convective flux, and usually a flux
divergence which leads to additional cooling. This is intended to mimic the
cooling effect by overshooting often observed in 3D models in the upper
photosphere.

%------------------------------------------------------------------------
\section{Results and discussion}
%------------------------------------------------------------------------

\subsection{Properties of surface granulation\label{s:granulation}}

\begin{figure}
\centering
\includegraphics[width=\columnwidth]{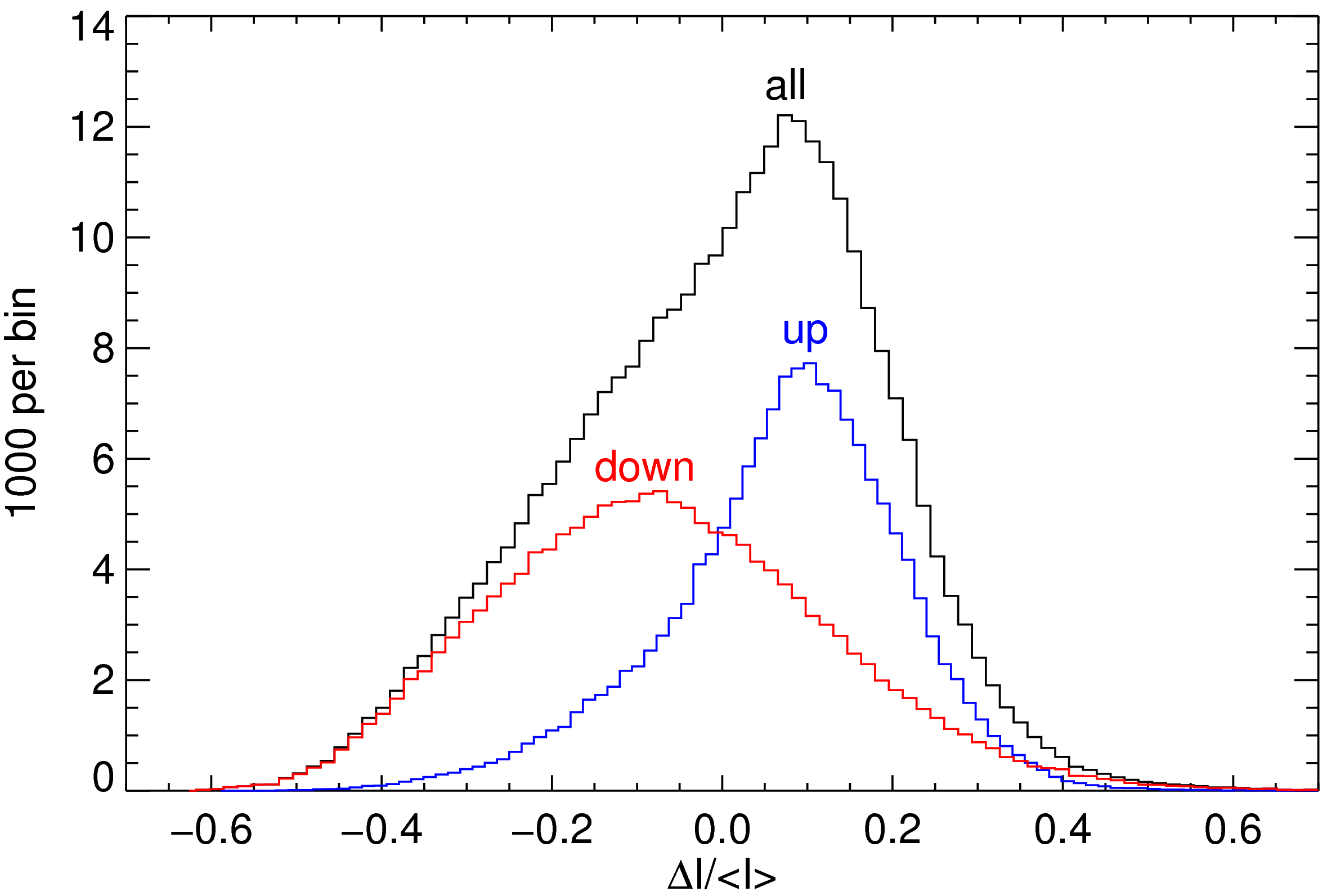}
\caption{Non-normalized probability density of the emergent white light
  intensity extracted from 14 snapshots of the 3D model. ``all'' labels the
  overall distribution, ``up'' and ``down'' are the distributions restricted
  to up- or down-flowing material only. Intensities are given as relative
  deviation from the (temporal and spatial) mean intensity. As velocity
  criterion the sign of the vertical velocity component at
  $\log\tau_\mathrm{ross}=0$ was taken.}
\label{f:intenspdf}
\end{figure}

The 3D model of red giant predicts the existence of surface granulation which
is clearly visible in the time series of emergent white light intensities
shown in Fig.~\ref{f:timeseries}. Although similar granulation patterns are
routinely seen in the 3D models of the Sun \citep{N82}, M-dwarfs
\citep{LAH02}, white dwarfs \citep{FLS96}, brown dwarfs \citep{FAL10},
pre-main sequence stars \citep{LAH06}, subgiants \citep{LAH06} and warm giants
\citep{CAT07}, the existence of granulation in cool red giants is not exactly
self-evident: MLT predicts that the surface convective zone is confined to
optically thick layers in the 1D models of giants, with an only thin,
marginally unstable region extending into the upper atmosphere (see the 1D
stratification of the convective velocity in Fig.~\ref{f:vrms}). Consequently,
the outer optically thin regions are essentially unaffected by convection in
1D models.

%This feature is not particular to the \LHD\ model used in this work but generic to all 1D model atmospheres of red giants.

In the 3D model, the geometric distance between the upper boundary of the
convective region and the optically thin region is, however, not large. The
material, therefore, crosses the formal boundary of the convective region and
overshoots into the upper atmospheric layers. Similarly to dwarfs, the
efficiency of convective overshoot exhibits an exponential decline with height,
e.g., as hinted at by the shape of the
vertical velocity profile in the optical depth range {\mbox {$\log
    \tau_{\rm Ross}\sim1.0\dots -2.0$}} (Fig.~\ref{f:vrms}). According to
\citet{FLS96}, such exponential decline is caused by the penetration of
convective modes with long horizontal wavelengths into the formally
convectively stable layers.

It may perhaps be interesting to note that intergranular lanes are narrower in
our giant model than those seen in the 3D models of the Sun but wider than in
the models of late-type dwarfs \citep[cf.][]{LAH06}. Why this is so is not
immediately clear, although the increase of the relative width of down-drafts
with \Teff\ suggests that it may be related to a stronger smoothing of thermal
inhomogeneities caused by the more intense radiative energy exchange at higher
temperatures.

\begin{figure}
\centering
\includegraphics[width=\columnwidth]{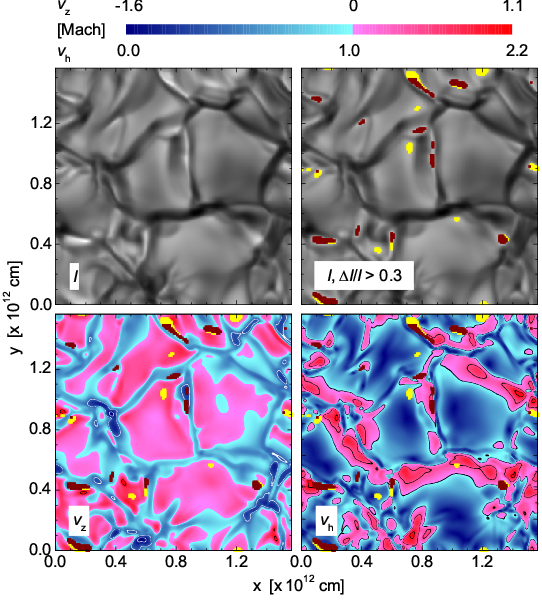}
\caption{Emerging white light intensity (top panels), vertical ($v_{\rm z}$)
  and horizontal ($v_{\rm h}=\sqrt{v_{\rm x}^2 + v_{\rm y}^2}$) velocities
  (bottom panels) in a typical 3D snapshot of the red giant model, all at
  $\log \tau_{\rm Ross}=0$. Yellow (bright) and brown (dark) pixels in the
  top-right and bottom panels highlight the areas where the relative intensity
  deviation is larger than $\Delta I/\langle I\rangle=0.3$ in the up-flows and
  down-flows, respectively. Locations where the vertical velocity $v_{\rm
    z}<-1$\,Mach (supersonic down-flows) are marked with white contours, those
  where the vertical and horizontal velocities exceed 1\,Mach are contoured in
  black (contours start at ${\rm Mach}=\pm1.0$ and continue at $\pm0.5$\,Mach
  thereof). The relative white light intensity contrast is 18.2\,\% for this
  particular snapshot.}
\label{f:intensmaps}
\end{figure}

Despite the fact that the granulation pattern shown in Fig.~\ref{f:timeseries}
exhibits the typical appearance known from the 3D model atmospheres of other
types of stars, Fig.~\ref{f:intenspdf} illustrates that its intensity
distribution does not show the familiar bimodal shape related to the dark and
bright areas of the granulation pattern but a single maximum. The intensity
distributions restricted to up- and down-flowing regions makes it clear that
the correlation between hot up-flowing and cool down-flowing gas is still
present, a clear indication that convection as such takes place.

However, the intensity distributions in Fig.~\ref{f:intenspdf}
illustrate that this correlation is far from perfect. For instance, high
surface brightness does not necessarily imply that matter is in the up-flow
since some of the down-flows may also be very bright. As seen in
Fig.~\ref{f:intensmaps}, such bright down-flows are typically seen on the
edges of convective cells and in many cases are located immediately next to
the uprising material that is also characterized by high white light
intensities. The high intensity in these regions is caused by
their significantly higher temperatures (see Fig.~\ref{f:tmaps}). These are,
in turn, caused by the dissipation of fast horizontal flows (mostly weak shock
waves) when they collide with the down-flowing material at the granule edges
are and are deflected downwards. This also explains why in some cases (but not
always) the brightest down-flows are located immediately next to the brightest
up-flows. There are also isolated regions of brightest intensity that are
located within the granules and which are tracing the hottest uprising
material.

Similar edge-brightened granules are seen on the surface of the Sun too, both
in observations \citep[e.g.,][]{KV92} and 3D models
\citep[e.g.,][]{SN98}. However, horizontal and vertical velocities are
significantly higher in the atmosphere of the red giant model. While only mild
shocks are seen in the models of the Sun, with the maximum Mach numbers of
$\sim$\,1.5 and $\sim$\,1.8 in the vertical and horizontal directions,
respectively, the corresponding numbers in the outer atmosphere of red giant
may reach to $\sim$\,2.5 and $\sim$\,6.0 (Fig.~\ref{f:v3maps}). It should also
be mentioned that some of the brightest spots in the up-flows sometimes appear
not on the edges but closer to granule centers, delineating the regions that
are splitting into new granules (Fig.~\ref{f:intensmaps}).

The average intensity contrast of the granular pattern seen in our model red
giant is $\sim$\,18.1\% (calculated for a sequence of 70 3D snapshots). This
is somewhat higher than the intensity contrast in the 3D atmosphere models of
the Sun ($\sim$\,15\%) but slightly lower than the corresponding number in a
model of a subgiant \citep[$\sim$\,23\%,][]{LAH02}.

The size of a typical granule in the red giant model is of the order of
$\sim$\,5\,Gm (Fig.~\ref{f:timeseries}), \change{somewhat larger than 10 times the
pressure scale height at the surface -- depending on the exact location and
whether turbulent pressure is considered in its definition or not. The
relative size is on the high side in comparison to models of higher gravity. 
The high horizontal velocities found in the model are in part a consequence of
this large granular size.
In absolute terms, the size}
is by
at least three orders of magnitude larger than the typical size of solar
granules, roughly equal to $\sim$\,1\,Mm \citep[e.g.,][]{NSA09}. The latter
number translates into $\sim2\times10^6$ granules on the Sun, in contrast
to only $\sim$\,400 granules on the surface of the red giant studied
here. This, together with the high intensity contrast, indicates that
granulation causes larger temporal fluctuations of the observable properties of
giants than of dwarfs, in the simplest case of their brightness.

\subsection{Thermal structure of the atmosphere}

\subsubsection{Properties of the full 3D structures\label{s:3Dstruct}}

\begin{figure}
\centering
\includegraphics[width=8.1cm]{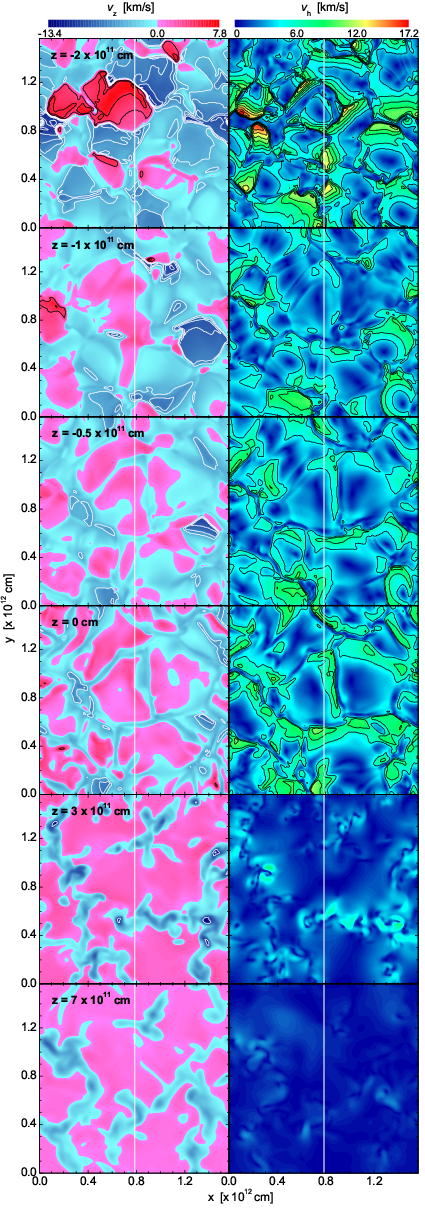}
\caption{Vertical (left) and horizontal (right) velocity maps in a typical
  3D snapshot of the red giant model constructed at different geometrical
  depth, $z$, with $z=0$ set at $\log \tau_{\rm Ross}=0$ and increasing
  towards the stellar center. Locations where the vertical velocity $v_{\rm
    z}\leqslant-1$\,Mach (supersonic down-flows) are marked with white
  contours, those where $v_{\rm z}$ and $v_{\rm h}$ exceed 1\,Mach are
  contoured in black (contours start at ${\rm Mach}=\pm1.0$ and are drawn at
  every $\pm0.5$\,Mach thereof). Vertical lines mark the $x$ position
  ($x=7.77\times10^{11}$\,cm) at which the 1-dimensional velocity profiles
  shown in Fig.~\ref{f:v3profiles} were taken.}
\label{f:v3maps}
\end{figure}

\begin{figure}
\centering
\includegraphics[width=8.1cm]{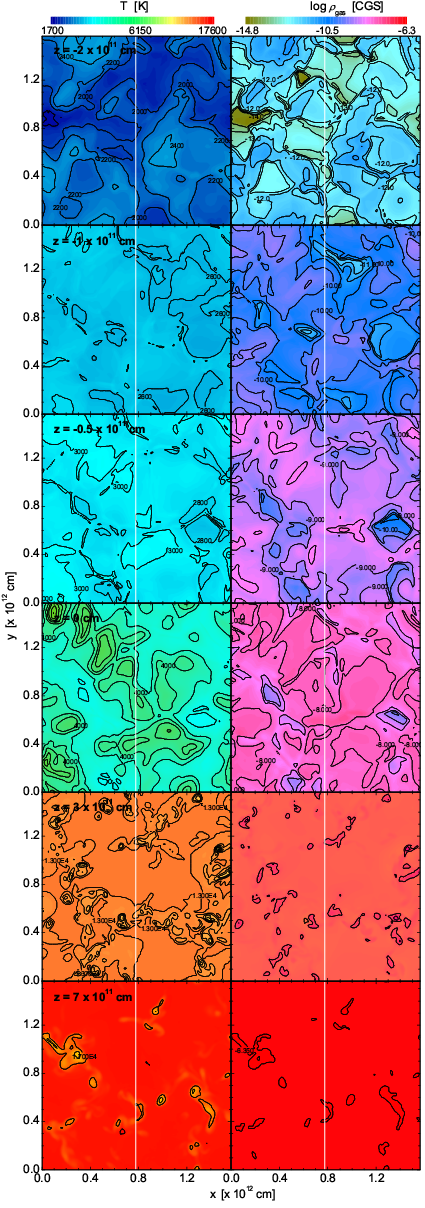}
\caption{Temperature (left) and density (right) maps in the typical 3D
  snapshot of the red giant model constructed at different geometrical depth,
  $z$, with $z=0$ set at $\log \tau_{\rm Ross}=0$ and increasing towards the
  stellar center. Contours are drawn at every 200\,K (panels 1--3, top-down)
  and 500\,K (panels 4--6) for temperature, and 0.5\,dex (panels 1--4,
  top-down) and 0.1\,dex (panels 5--6) for density. Vertical lines mark the
  $x$ position ($x=7.77\times10^{11}$\,cm) at which the 1-dimensional velocity
  profiles shown in Fig.~\ref{f:tprofiles} were constructed.}
\label{f:tmaps}
\end{figure}

As it was already discussed in Sect.~\ref{s:granulation}, the 3D model shows a
prominent granulation pattern, a direct consequence of convective
motions. Convective cells are discernible in the deeper (and hotter)
atmospheric layers and are pronounced between the optical surface (at
$z=0$\,cm or \mbox{$\langle\tau_{\rm Ross}\rangle=1$}) and the lower boundary
of the model at $z=7\times10^{11}$\,cm (Figs.~\ref{f:v3maps} and
\ref{f:tmaps}). In this depth range convection manifests itself in the form of
wide up-flows and narrower and cooler down-flows.

Velocity amplitudes increase in the higher atmospheric layers and their
vertical velocity, $v_{z}$, sometimes becomes supersonic (upper panels in the
left column of Fig.~\ref{f:v3maps}). The granulation pattern looses coherence
towards the upper atmosphere, and the velocities become gradually dominated
by motions related to acoustic waves.

\begin{figure}
\centering
\includegraphics[width=7.7cm]{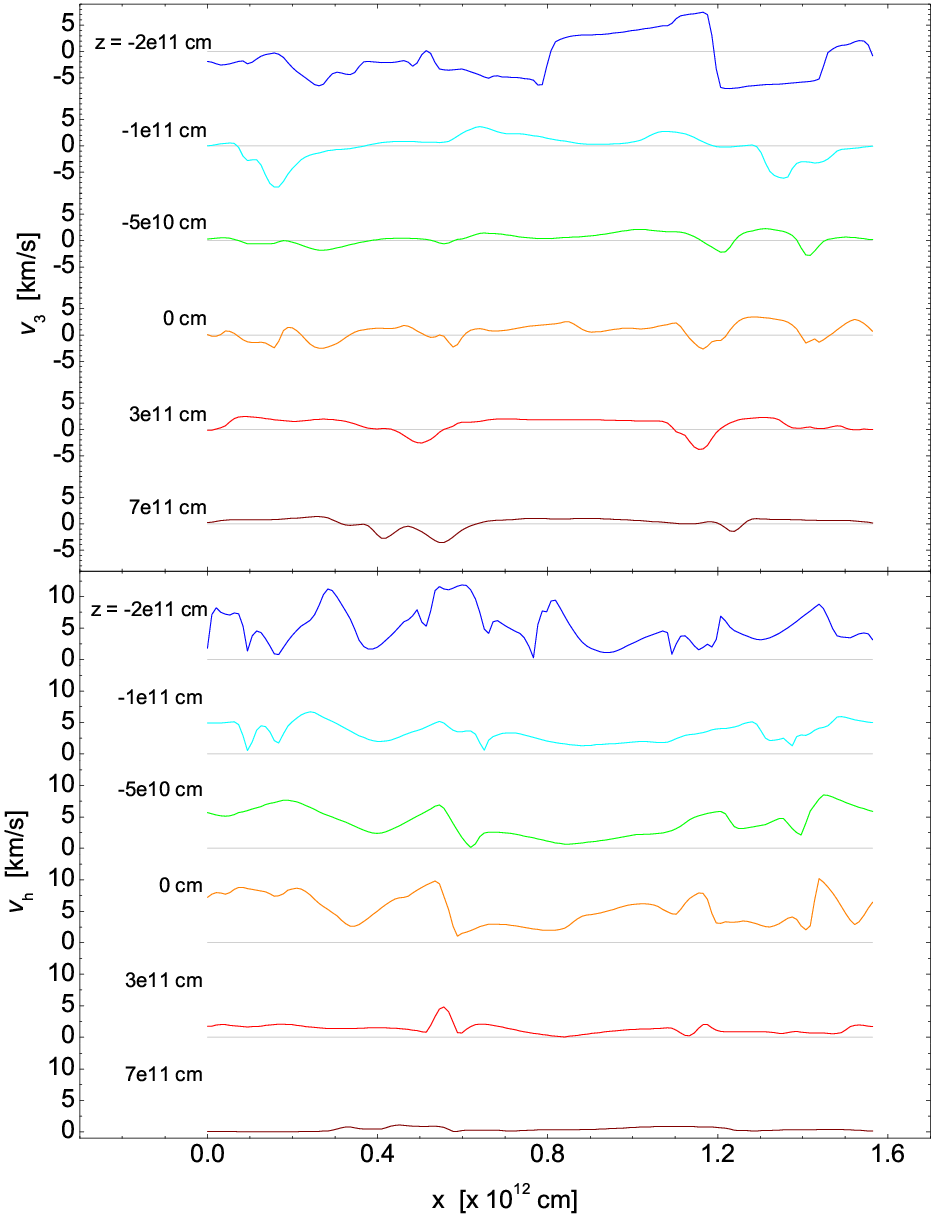}
\caption{1-dimensional vertical (top) and horizontal (bottom) velocity
  profiles in the red giant model at $x=7.77\times10^{11}$\,cm (marked the by
  white vertical lines in Fig.~\ref{f:v3maps}) and shown at different
  geometrical depths as indicated on the left-hand side of each velocity
  profile. Grey horizontal lines indicate the level where the velocity is equal to
  zero.}
\label{f:v3cut}
\end{figure}

\begin{figure}
\centering
\includegraphics[width=8.3cm]{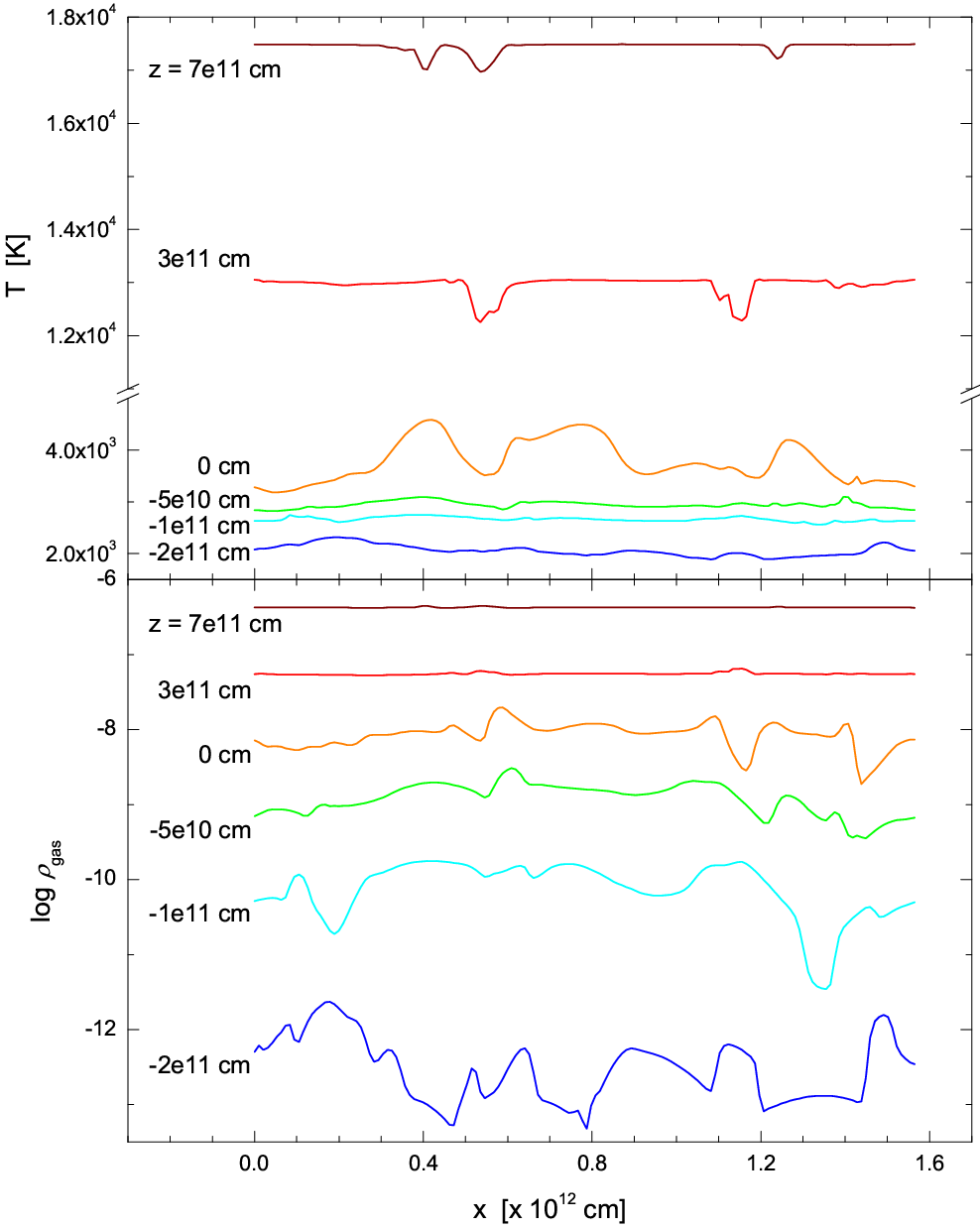}
\caption{1-dimensional profiles of temperature (top) and density (bottom) at
  $x=7.77\times10^{11}$\,cm (marked by the white vertical lines in
  Fig.~\ref{f:tmaps}) and shown at different geometrical depths.}
\label{f:tcut}
\end{figure}

\begin{figure}
\centering
\includegraphics[width=\columnwidth]{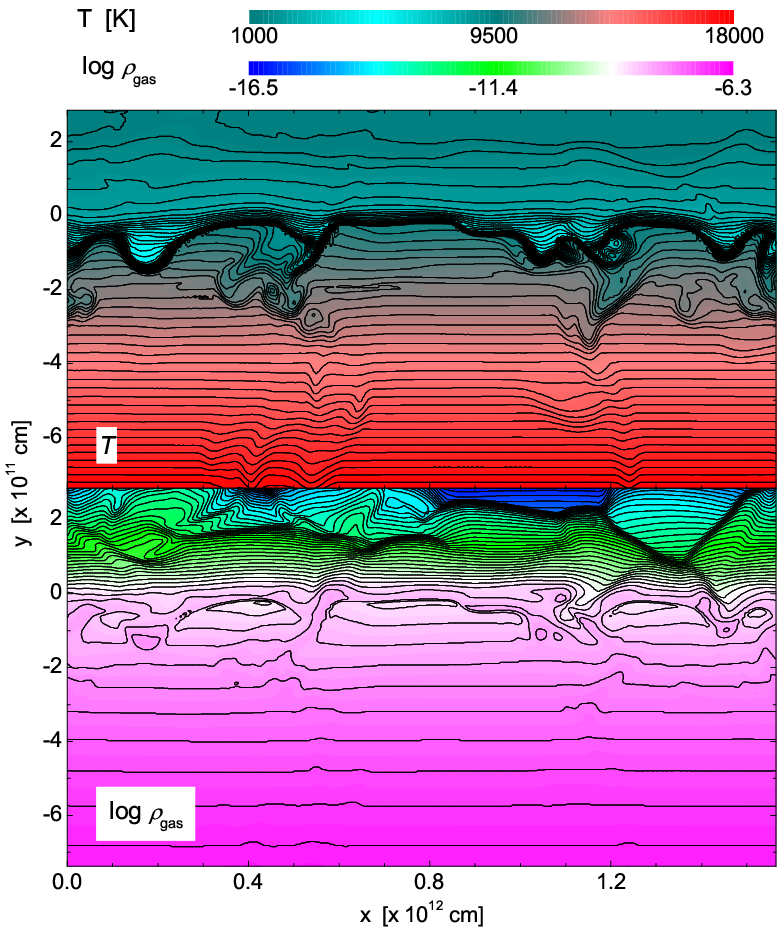}
\caption{2D profiles of temperature (top) and density (bottom) in the red
  giant model at $x=7.77\times10^{11}$\,cm (marked by white vertical lines in
  Fig.~\ref{f:tmaps}). Contours are drawn at every 250\,K (top) and 0.2\,dex
  (bottom).}
\label{f:tprofiles}
\end{figure}

The horizontal flow speed, $v_{\rm h}$, is low in the subphotospheric layers
due to the predominantly vertical motions here. The situations starts to
change when matter approaches the optical surface. The decreasing opacity
leads to enhanced photon losses, the matter rapidly cools, becomes denser, and
its vertical velocity decreases. A density inversion is formed just below the
optical surface, at $\log\tau_{\rm Ross}\approx0.5$ (see
Fig.~\ref{f:tprofiles}). Simultaneously, the up-flowing material is
deflected sideways until it reaches the granule edges finally merging into the
down-flows. The resulting pattern of horizontal velocities is tracing the
granular shapes and is visible in the atmospheric layers above the optical
surface (Fig.~\ref{f:v3maps}).

Figure~\ref{f:v3cut} illustrates that the amplitude of the velocity fluctuations
on horizontal planes is highest in the outermost layers, and largely shaped by
shock waves. This holds for both the vertical and horizontal
velocities. Towards deeper layers, the fluctuations in the vertical velocity
become noticeably smaller, fluctuations of the horizontal velocity also
decrease but to a lesser degree. As alluded to before, this is a consequence
of the take-over of the more regular convective motions over wave motions.
After passing a minimum around the optical surface the fluctuations of the
vertical velocity increase again, a signature of the roughly columnar flow
pattern in the the subphotospheric layers.

\begin{figure}
\centering
\includegraphics[width=\columnwidth]{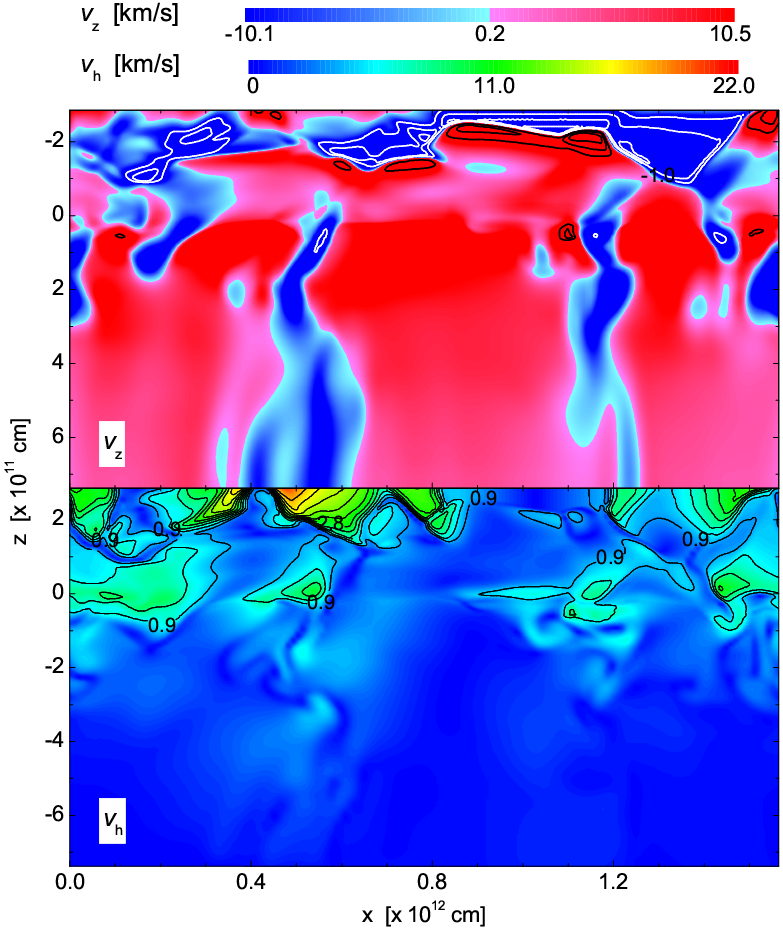}
\caption{2D profiles of vertical (top) and horizontal (bottom) velocity in the
  red giant model at $x=7.77\times10^{11}$\,cm (marked by white vertical lines
  in Fig.~\ref{f:v3maps}). Locations where the vertical velocity $v_{\rm
    z}\leqslant-1$\,Mach (supersonic down-flows, top panel) are marked with
  white contours, those where $v_{\rm z}$ and $v_{\rm h}$ exceed 1\,Mach are
  contoured in black (contours start at ${\rm Mach}=\pm1.0$ and are drawn at
  every $\pm0.5$\,Mach thereof).}
\label{f:v3profiles}
\end{figure}

As shown in Fig.~\ref{f:tcut} the columnar pattern in the subphotosphere is
also imprinted in the temperature field which is rather homogeneous in the
up-flows, and drops in the temperature marking down-flows. The temperature
fluctuations reach their maximum in the surface layers, and are quickly
reduced in amplitude in the optical thin layers. This is surprising in view of
the substantial density fluctuations (see Fig.~\ref{f:tcut}, lower panel)
present in the photospheric layers. This indicates an efficient smoothing of
the temperature field by radiative energy exchange counteracting temperature
changes by adiabatic compression or expansion.  The vertical cuts in
Fig.~\ref{f:v3profiles} and \ref{f:tprofiles} show that the shock waves form a
rather irregular pattern in the upper atmosphere. The shock fronts are often
horizontal or arc-like, similar to those seen in the 3D hydrodynamical models
of the Sun \citep[e.g., ][]{WFSLH04}. Most of the shock activity takes place
above $z=-0.5\times10^{11}$\,cm ($\log \tau_{\rm Ross}<-1.0$) where the flow
density is low enough for the flow to accelerate to supersonic speeds.

Finally, we would like to point out that our model is rather shallow in the
sense that its extent below the optical surface supercedes the horizontal size
of granular cells only little. Consequently, the merging and narrowing of
downdrafts with depth as discussed by \citet{SN98} is not clearly
discernable. However, qualitative similarity of the convective morphology
makes us belief that models with larger extent in depth will exhibit the same
feature.

\subsubsection{Properties of the \xtmean{\mbox{3D}} stratification}
\label{s:propaver}

\begin{figure}
\centering \includegraphics[width=\columnwidth]{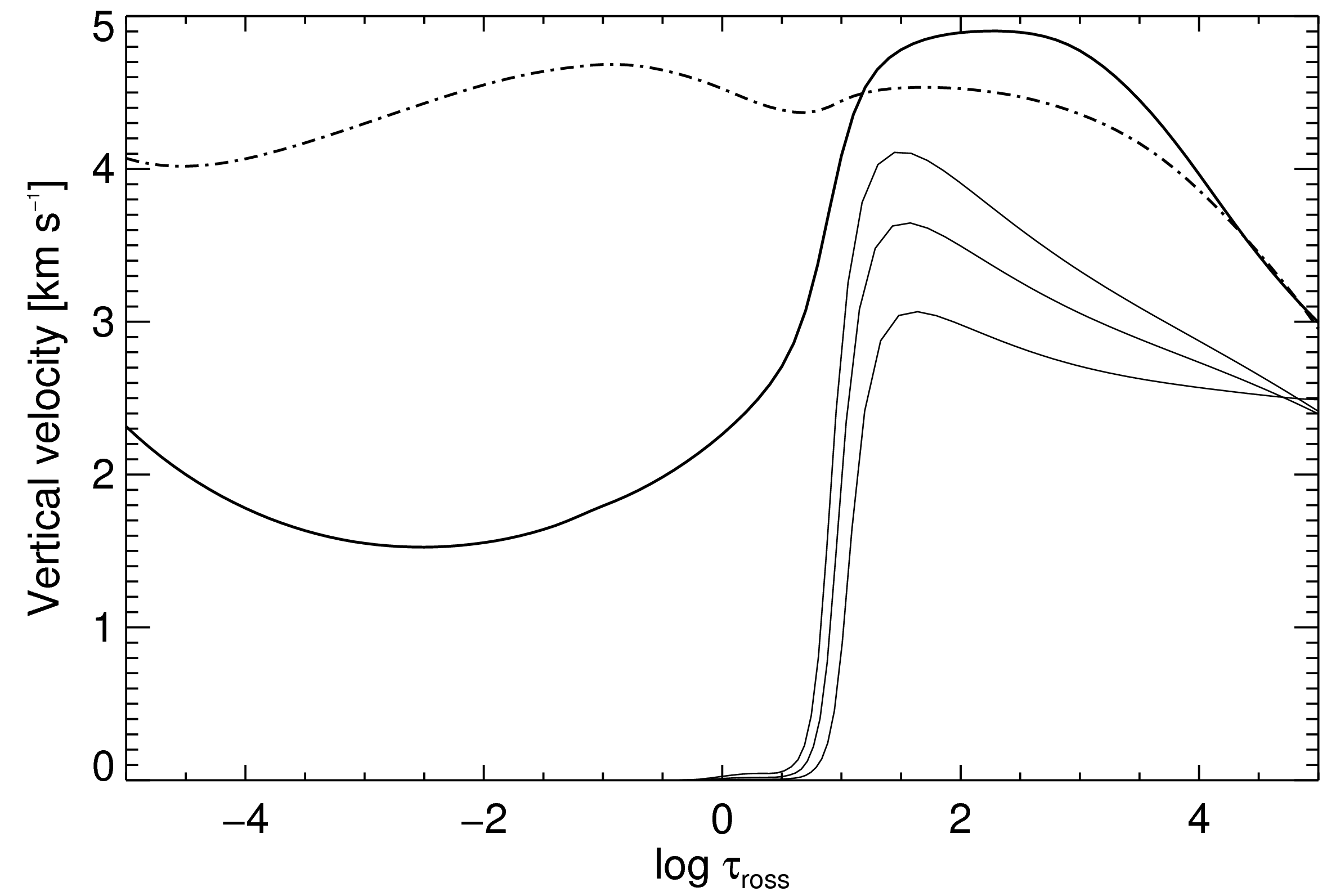}
\caption{Temporally and horizontally averaged RMS vertical velocity (thick
  solid line) and RMS horizontal velocity (thick dot-dashed line) of the 3D
  hydrodynamical model in comparison to the 1D \LHD\ models (thin lines) as a
  function of Rosseland optical depth. The \LHD\ models were calculated using
  different mixing-length parameter, $\mlp=1.0,1.5,2.0$ (top-down).}
\label{f:vrms}
\end{figure}

\begin{figure}
\centering
\includegraphics[width=\columnwidth]{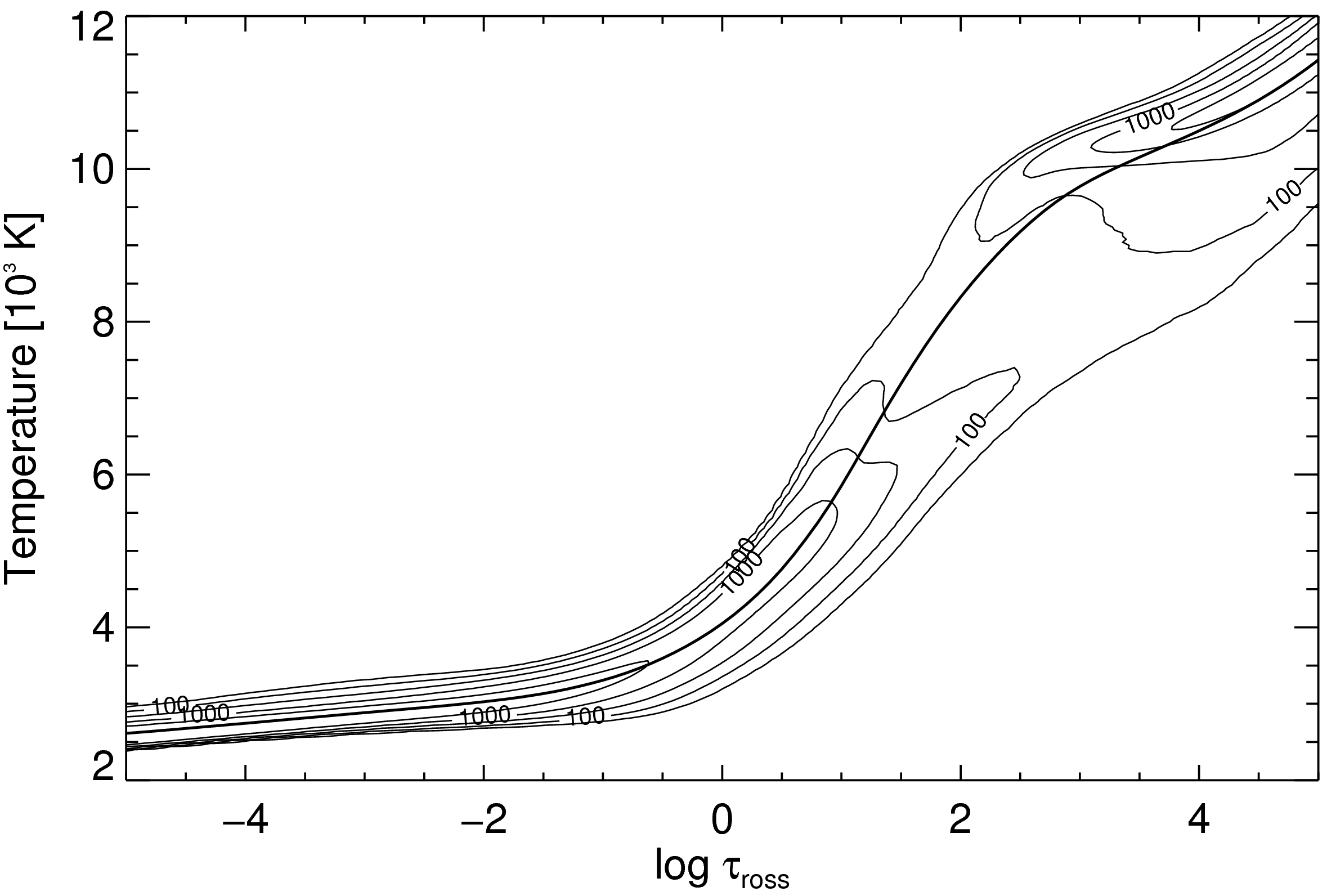}
\caption{Non-normalized joint probability density function of temperature and
  optical depth (thin contour lines), and average temperature profile (thick
  solid line) of the 3D model. The statistics was obtained from three
  snapshots of the flow field. The contour lines are plotted in steps of
  factors of $\sqrt{10}$.}
\label{f:ttaupdf}
\end{figure}

\begin{figure}
\centering
\includegraphics[width=\columnwidth]{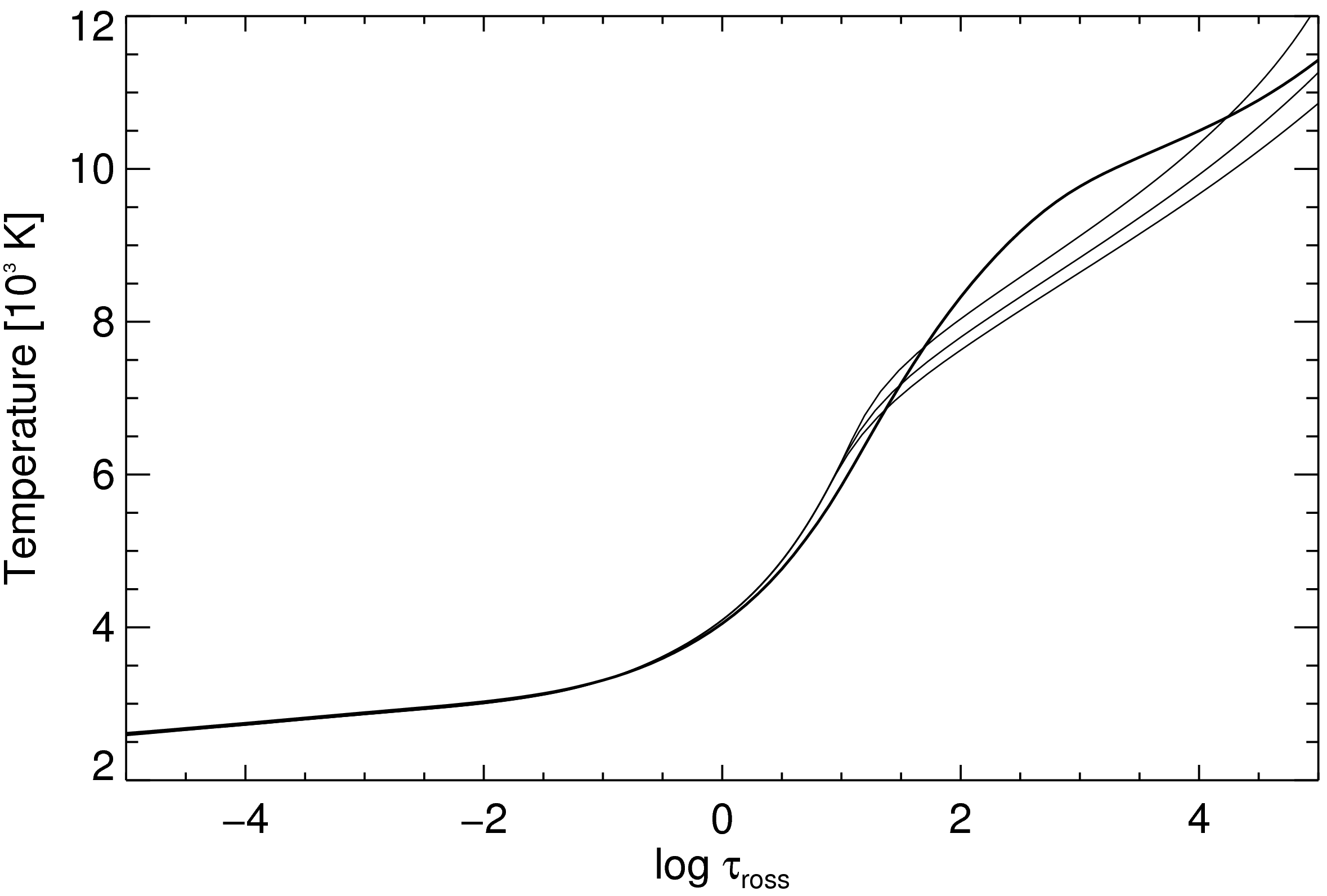}
\caption{Temperature profile of the \xtmean{\mbox{3D}} model (thick solid
  line) in comparison to 1D \LHD\ models (thin solid lines) as a function
  of Rosseland optical depth. The \LHD\ models were calculated with different
  mixing-length parameters, $\mlp=1.0,1.5,2.0$ (upper$\rightarrow$lower).}
\label{f:ttau}
\end{figure}

As we have already seen in Sect.~\ref{s:3Dstruct}, two regions can be
distinguished in the photosphere of the giant model: the lower (and
hotter) part dominated by convective motions {\it per se} and upper layers
where the flow is driven by acoustic waves. This substructure clearly
manifests itself in the (temporally and horizontally averaged) RMS vertical and
horizontal velocity profiles (Fig.~\ref{f:vrms}). Convective motions gradually
cease beyond the formal convective boundary at $\log \tau_{\rm Ross}\sim
0.5$. However, the matter partially penetrates beyond it in a form of
exponentially decreasing overshoot. Qualitatively, this is very similar to the
picture seen in late-type dwarfs \citep{LAH02,LAH06}.

The temperature profile of the
\xtmean{\mbox{3D}} model follows closely the ridgeline corresponding to the
maxima of the probability density distribution of temperature and optical
depth in the 3D hydrodynamical model (Fig.~\ref{f:ttaupdf}). A sharp decrease
both in the 3D and \xtmean{\mbox{3D}} temperature profiles between the optical
depths of $\log \tau_{\rm Ross}=0$ and 2.0 is caused by the rapid increase in
the radiative cooling rate close to the optical surface.

Interestingly, the temperature profile corresponding to the \xtmean{\mbox{3D}}
model is markedly different from the 1D temperature profiles in the deeper
atmosphere, at $\log \tau_{\rm Ross}>0.5$ (Fig.~\ref{f:ttau}). Moreover, none
of the 1D \LHD\ models with different mixing-length parameters, $\alpha_{\rm
  MLT}$, is able to satisfactorily reproduce the stratification of the
\xtmean{\mbox{3D}} model. The resulting different pressure--temperature
relation have a direct influence on the radius of the star. Note, that
turbulent pressure was neglected in the 1D LHD models at this stage.

\begin{figure}
\centering
\includegraphics[width=\columnwidth]{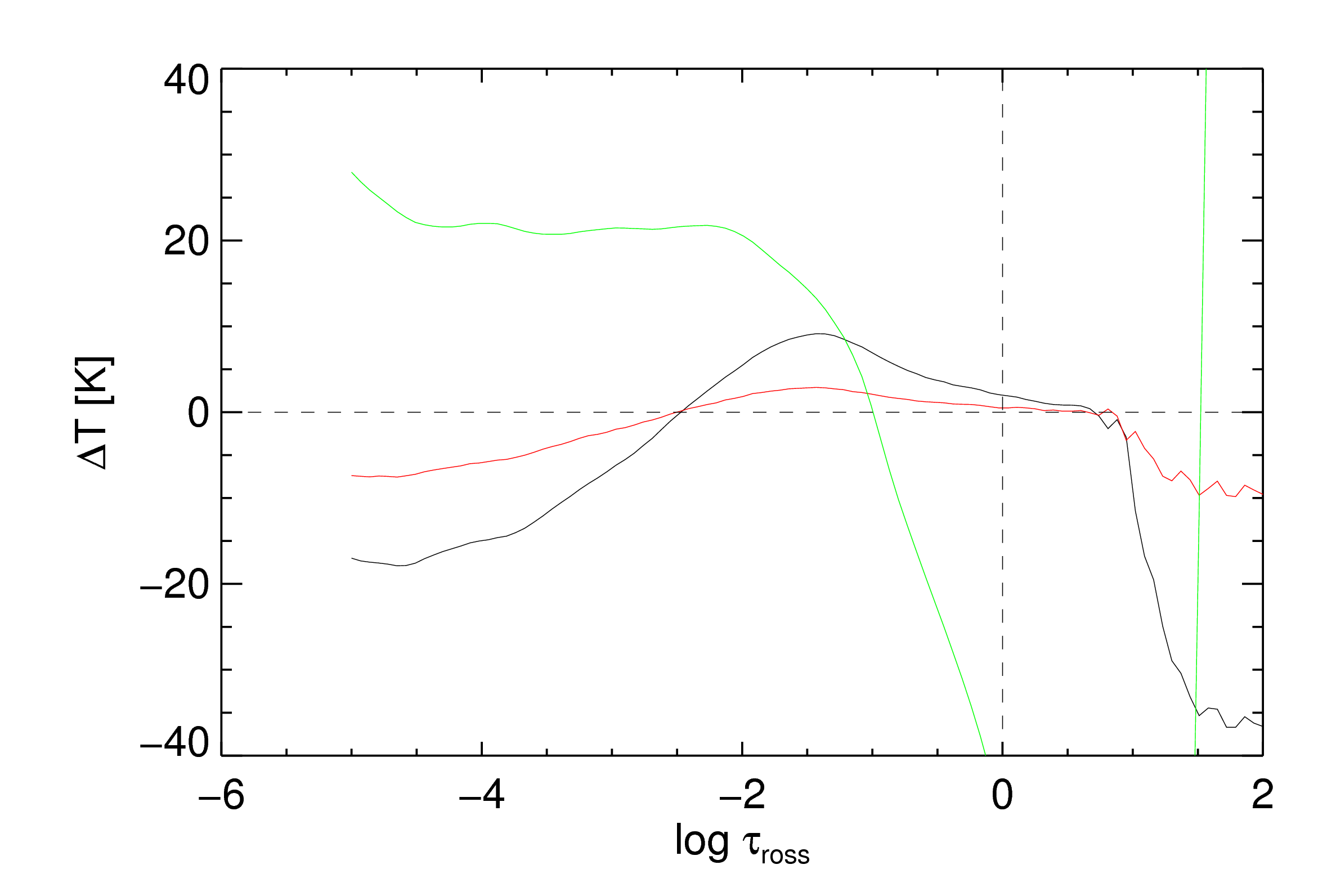}
\caption{Deviations of temperature profiles from the \LHD\ (\mlp=2.0)
  convective-radiative equilibrium model. The green line shows the
  \xtmean{\mbox{3D}} model, red and black lines depict
  \LHD\ models with different overshooting velocities, \vover=1\pun{km
    s$^{-1}$} and \vover=2\pun{km s$^{-1}$}, respectively.}
\label{f:dt}
\end{figure}

Figure~\ref{f:dt} shows that in the higher atmosphere the 3D model exhibits a
temperature \textit{increase} of about 20\pun{K} relative to pure radiative
equilibrium conditions. Energy fluxes show that this is not the consequence of
mechanical heating (e.g. by shocks). Only the radiative flux provides
significant energy exchange in these layers. \LHD\ models with an ad-hoc
overshooting velocity show a certain degree of cooling. The cooling is not
great despite the substantial overshooting velocities put into the 1D
models. This again illustrates the rather tight coupling of the temperature to
radiative equilibrium conditions as already seen by the rather small
horizontal temperature fluctuations in the upper photosphere.  We argue that
the heating in the 3D case is due to an altered radiative equilibrium
temperature in the presence of horizontal $T$-inhomogeneities and the specific
wavelength-dependence of the opacity (see
Appendix~\ref{appendixb}). Interestingly, this is opposite to what is seen in
the outer atmosphere of red giants at lower metallicities, where the average
temperature of the hydrodynamical model is significantly lower than that of
the corresponding 1D model \citep[e.g.,][]{CAT07,DKL10,IKL10}.

\subsubsection{Radiative, hydrodynamical and rotational time scales\label{s:timescales}}

\begin{figure}
\centering
\includegraphics[width=\columnwidth]{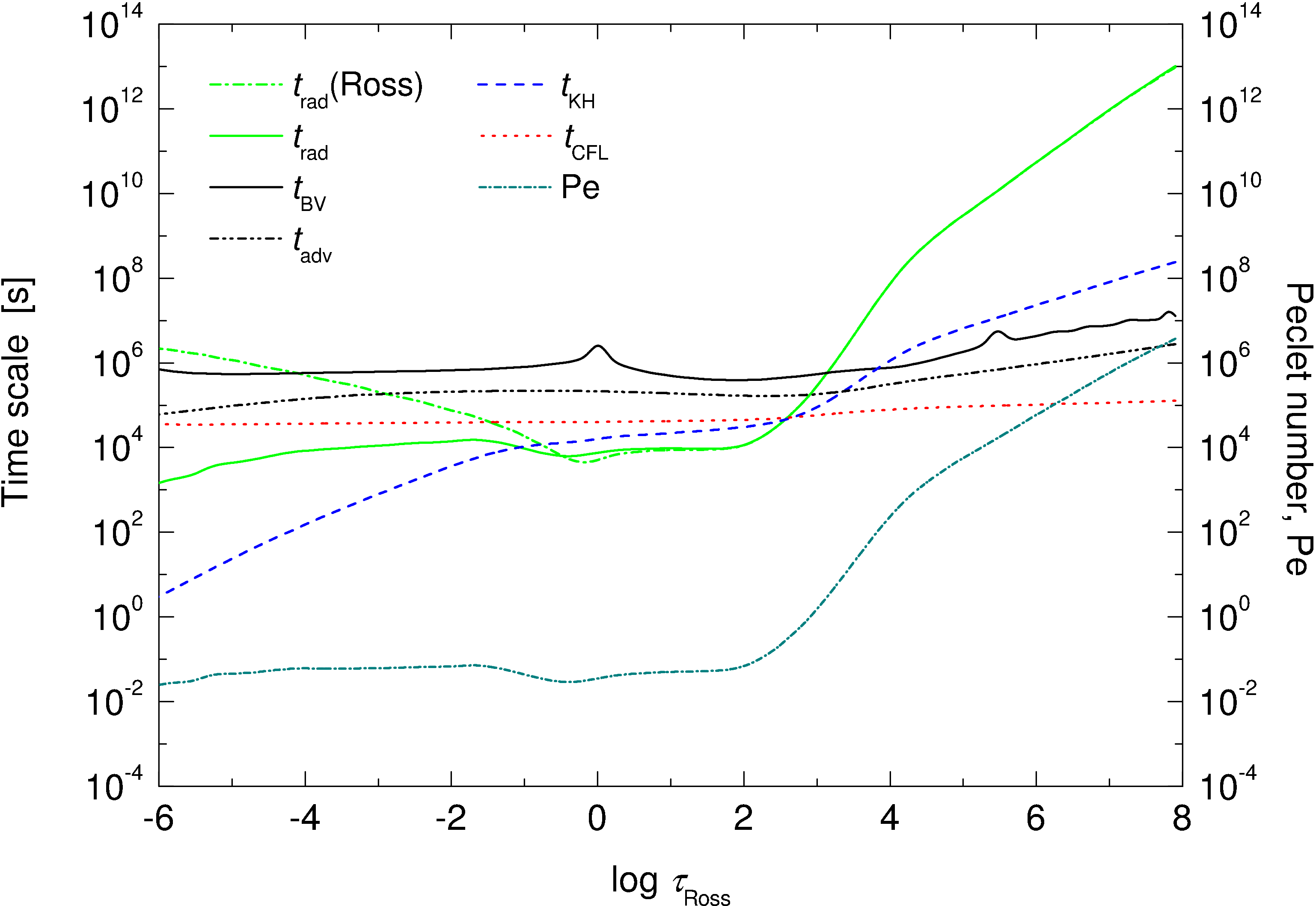}
\caption{Radiative and hydrodynamical time scales in the 3D atmosphere model
  of red giant.}
\label{f:timescales}
\end{figure}

In order to asses the possible interplay between various physical phenomena
that take place in the red giant atmospheres we calculated a number of
characteristic radiative and hydrodynamic time scales (see
Appendix~\ref{appendixa} for the definitions). The time scales were calculated
using the 3D model as background, and are plotted versus the optical depth in
Fig.~\ref{f:timescales}.

Two convection-related time scales, as given by the Brunt-V\"{a}is\"{a}l\"{a}
period, $t_{\rm BV}$, and the time for crossing one mixing-length, $t_{\rm
  adv}$, are similar within an order of magnitude and show little variation
throughout the entire atmosphere. The radiative time scale taking into account
the wavelength dependence of the opacity, $t_{\rm rad}$ is significantly
larger than convective times scales in the deep atmosphere where the flow is
nearly adiabatic. However, $t_{\rm rad}$ rapidly decreases with increasing
height and above $\log\tau_{\rm Ross}\sim2$ becomes about two orders of
magnitude shorter than either $t_{\rm BV}$ or $t_{\rm adv}$, thus causing a
rapid evolution of the flow towards radiative equilibrium. For further
comparison we added the radiative time scale $t_{\rm rad}\mbox{(Ross)}$ based
on grey Rosseland opacities. In the optically thin region it substantially
overestimates the radiative relaxation time.

The time scale of thermal relaxation, or Kelvin-Helmholtz time scale, $t_{\rm
  KH}$ is considerably larger than the convective time scale deep in the
atmosphere ($\log \tau_{\rm Ross}>4$). This may suggest that the flow in the
deeper atmosphere will not be able to relax to thermal equilibrium during the
model simulation time, $\sim 6\times10^6$\pun{sec} (see Sect.~\ref{s:model}),
which is $\sim1-2$ orders of magnitude smaller than $t_{\rm KH}$. However,
such conclusion would be wrong, since at these depths thermal relaxation is in
fact governed by the mass exchange due to convection which takes place on
significantly shorter time scales.

The assessment of whether stellar rotation may influence the dynamical and
radiative processes in the atmospheres of red giants is not a trivial task,
especially since their rotational periods are largely unknown. In order to
obtain a qualitative picture one may nevertheless compare the relevant time
scales. To derive an estimate of the rotational time scale we resorted to the
available measurements of projected rotational velocities, $v_{\rm rot} \sin
i$. For red giants these are are typically in the range of $2-9$\pun{km/s}
\citep[e.g.,][]{CGY08,CSRB09}. Ignoring the fact that rotational velocity,
$v_{\rm rot}$, may depend on the effective temperature \citep{CGY08} and
metallicity \citep{CSRB09}, and taking $3$\pun{km/s} as a representative
rotational velocity, together with a radius of {\mbox{$\sim\,75\,{\rm
      R_{\odot}}$}} (see Sect.~\ref{s:model}) one obtains the rotation period
of $\sim10^8$\pun{sec}\footnote{This compares well with, e.g., the rotational
  period of $2\pm0.2$\,yr ($\sim6.3\times10^7$\pun{sec}) derived by
  \citet{GB06} for Arcturus, the atmospheric parameters of which are not too
  different from the red giant studied here.}. This is by at least 1--2 orders
of magnitude larger than the convective and radiative time scales, except in
the deepest layers where the radiative time scale is longer. While we therefore
conclude that rotation should have a minor influence on governing the
atmospheric dynamics of this particular red giant, it should be noted that in
the deeper atmosphere rotational and convective time scales may become
comparable and thus the interaction between convection and rotation
interaction may become non-negligible.

\subsubsection{The role of turbulent pressure}

\begin{figure}
\centering
\includegraphics[width=\columnwidth]{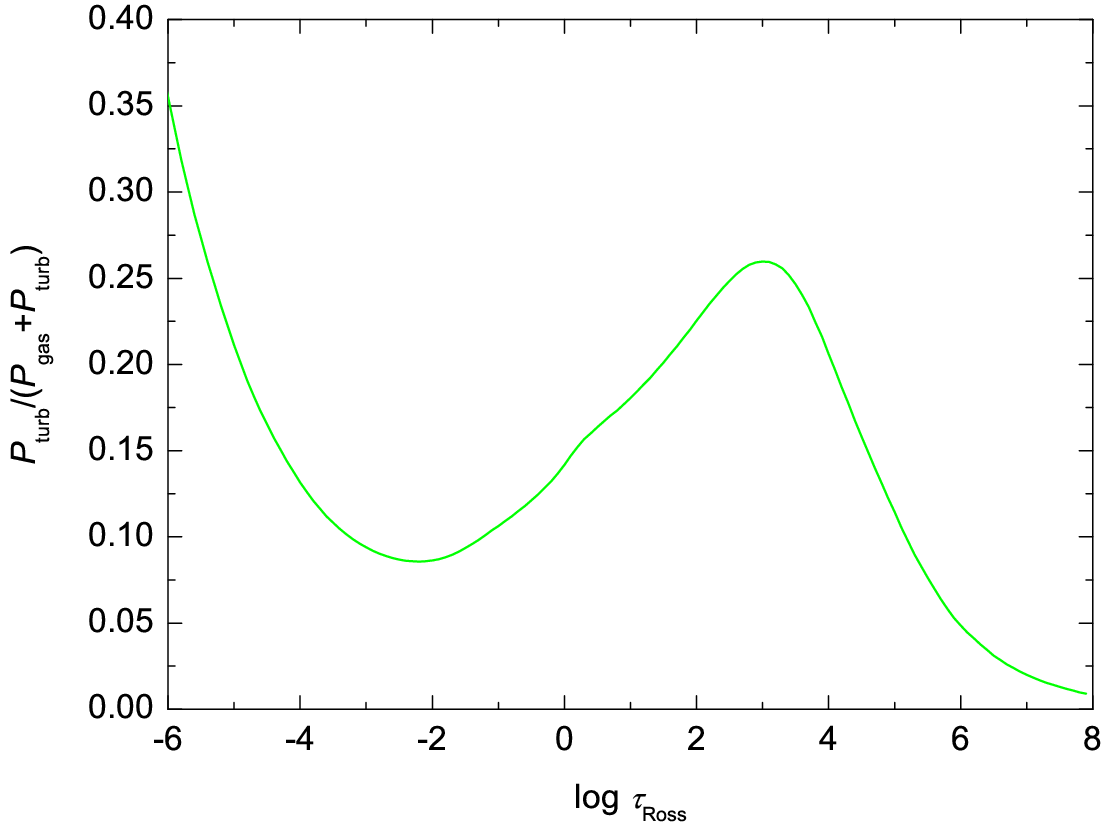}
\caption{Turbulent pressure to total pressure versus optical depth in the 3D model.\label{f:pturb}}
\end{figure}

One of fundamental differences between the 3D hydrodynamical and classical 1D
atmosphere models of red giants is that hydrodynamical models predict non-zero
turbulent pressure, $P_{\rm turb}$. The contribution of the turbulent pressure
to the total pressure (i.e., gas and turbulent) may be as large as $15\%$ in
the 3D hydrodynamical models of the Sun \citep{H10}. In order to assess its
importance in the red giant atmosphere we calculated the turbulent pressure as
$P_{\rm turb}=\langle\rho v_\mathrm{z}^{2}\rangle$, where $\rho$ and
$v_\mathrm{z}$ are gas density and velocity, respectively, and angular
brackets denote \change{temporal averaging as well as averaging} on surfaces of equal optical
depth. The ratio of turbulent to total pressure is plotted versus optical
depth in Fig.~\ref{f:pturb}.

\begin{figure}
\centering
\includegraphics[width=\columnwidth]{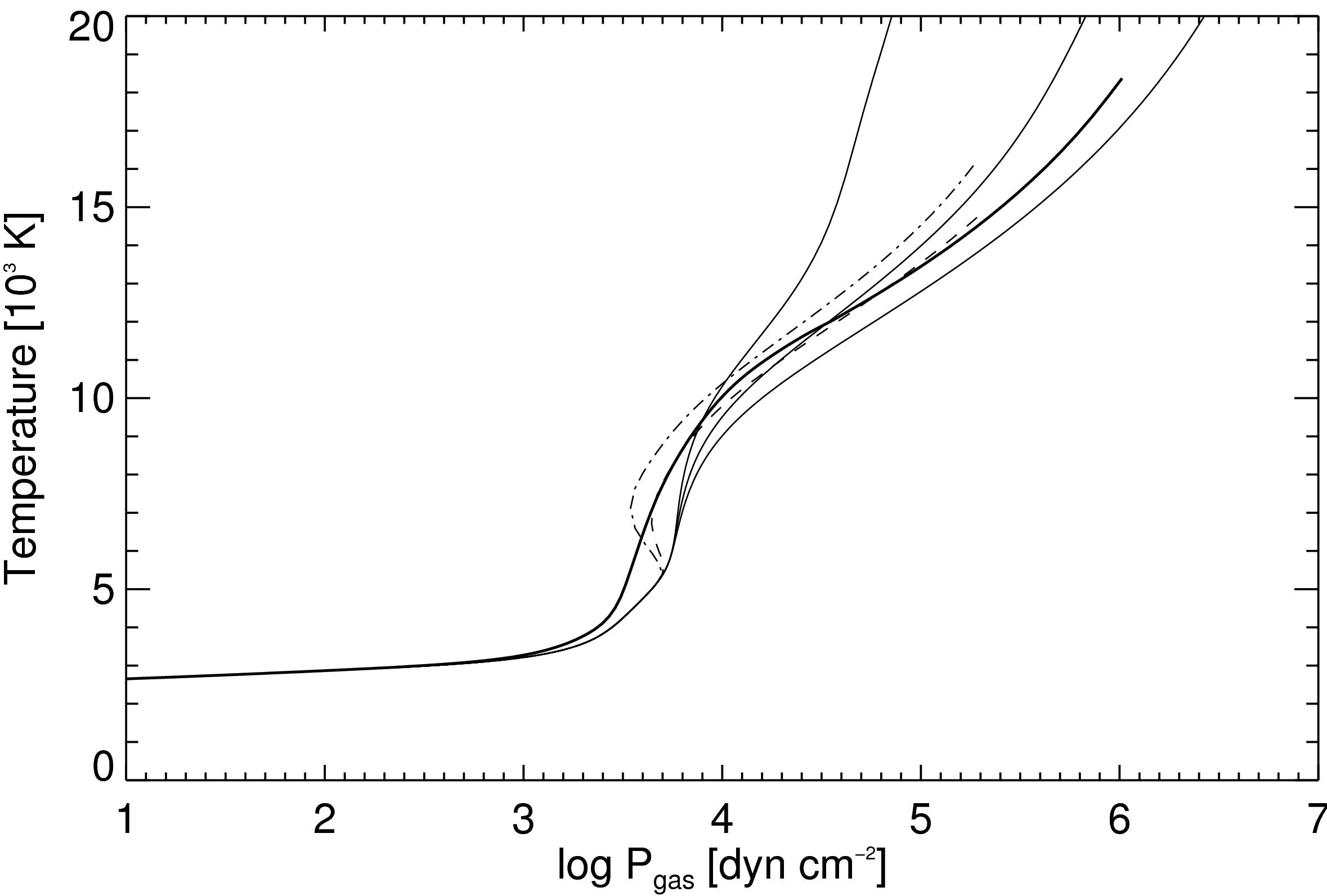}
\caption{Temperature stratification in the 3D hydrodynamical model of a red
  giant, as a function of gas pressure, $P_{\rm gas}$. The thick solid line is
  the profile of the \xtmean{\mbox{3D}} model,
  the thin lines are 1D LHD models calculated using different
  mixing-length parameter ($\alpha_{\rm MLT}=1.0,1.5,2.0$, from left to right;
  in all cases $P_{\rm turb}=0$). The two 1D models with non-zero turbulent
  pressure are shown by dashed ($\alpha_{\rm MLT}=2.0$ and $f=1.0$) and
  dashed-dotted ($\alpha_{\rm MLT}=2.0$ and $f=2.0$) lines.\label{f:tp}}
\end{figure}

Although rarely done in the routine calculations of, e.g., 1D stellar model
atmosphere grids, in principle turbulent pressure can be included in the
calculation of classical 1D atmosphere models, in addition to gas and electron
pressure. The result of such an exercise is shown in Fig.~\ref{f:tp}, where we
plot two \LHD\ models calculated with non-zero turbulent pressure ($P_{\rm
  turb}=f \rho v^{2}$, here $v$ is convective velocity as given in the
framework of MLT, and $f$ a dimensionless factor, usually $f<1$). Three
additional models shown there were constructed using different mixing-length
parameters, $\alpha_{\rm MLT}=1.0,1.5,2.0$, utilizing a formalism of
\citet{M78} and assuming a vanishing turbulent pressure, $P_{\rm turb}=0$. As
it is evident from Fig.~\ref{f:tp}, there is no reasonable combination of
$\alpha_{\rm MLT}$ and $f$ allowing to construct a classical 1D model which
would reproduce the pressure--temperature relation of 3D hydrodynamical model.

How important then is the contribution of turbulent pressure in the
atmospheres of red giants in general? In the 3D hydrodynamical model
atmosphere of a red giant studied here $P_{\rm turb}$ provides a non-vanishing
contribution to the total pressure nearly throughout the entire
atmosphere, except for the deepest layers where its influence is minor
(Fig.~\ref{f:pturb}). The distribution of $P_{\rm turb} / (P_{\rm gas} +
P_{\rm turb})$ is double-peaked: the first maximum at $\log \tau_{\rm
  Ross}\sim4$ is due to increasing relative importance of convective flow,
whereas the monotonous rise beyond $\log \tau_{\rm Ross}\sim-2$ is due to wave
activity in the upper atmosphere. The peak values are significantly larger
than those in the Sun \citep[cf.][]{H10}: the contribution of turbulent
pressure may amount to $\sim25\%$ and $\sim13\%$ due to convective motions
close to the optical surface in the red giant and the Sun, respectively.

Obviously, the importance of turbulent pressure can not be neglected in
realistic modeling of stellar atmospheres and stellar evolution. One may argue
that in the case of the red giant studied here the discrepancy between the
pressure--temperature relations of the 3D hydrodynamical and 1D models occurs
deep in the atmosphere ($\log \tau_{\rm Ross}\sim2$) and thus should have a
minor influence on the formation of emitted spectrum. However, this may be of
importance for giants at lower metallicities where convection can reach into
the layers beyond the optical surface. Additionally, larger pressure at any
given geometrical depth would lift the higher-lying stellar layers outwards
and may eventually change the stellar radius and luminosity, with direct
consequences on the evolutionary tracks and isochrones. Clearly, issues
related with the proper treatment of turbulent pressure in the current stellar
evolution and atmosphere modeling of red giants warrant further study.

\subsubsection{Effective mixing-length parameter for the 1D model atmospheres\label{s:mlp}}

\begin{figure}
\centering \includegraphics[width=\columnwidth]{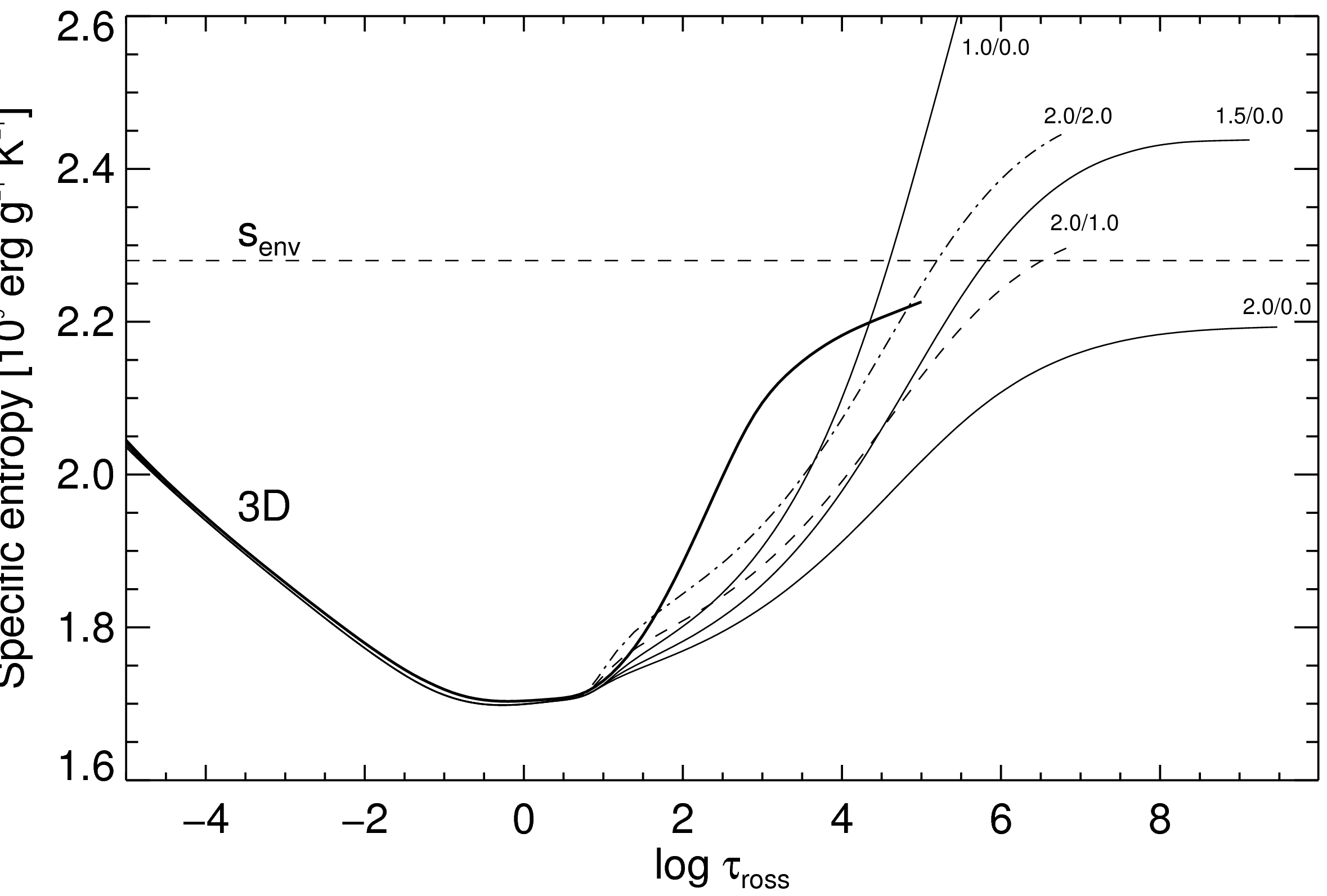}
\caption{Entropy profiles as a function of optical depth. The dashed line
  indicates the value of the asymptotically reached entropy in the 3D
  model. Thick solid line is entropy stratification of the temporally and
  spatially averaged \xtmean{\mbox{3D}} model, thin lines are 1D \LHD\ models
  calculated with different mixing-length parameters, $\mlp=1.0,1.5,2.0 $
  (left-to-right). The two models with non-zero turbulent pressure are shown
  by dashed ($\mlp=2.0$ and $f=1.0$) and dash-dotted ($\mlp=2.0$ and $f=2.0$)
  lines.}
\label{f:stau}
\end{figure}

The potential of the hydrodynamical model atmospheres for the calibration of
the mixing-length parameter, \mlp, was explored by \citet{LFS99}. The
authors have shown that the entropy profiles in the 2D models of solar-type
stars posses a minimum in the convectively unstable region. While the shape
and location of this entropy minimum is somewhat different in individual
spatially resolved profiles, the common property is that the entropy gradient
becomes very small in the deep atmospheric layers where the adiabatic up-flows
dominate. Such asymptotic behavior is also characteristic to 1D model
atmospheres but the asymptotic entropy values reached deep in the atmosphere
are different for different mixing-length parameters, \mlp. Therefore, the
height of the asymptotic entropy plateau in the 2D hydrodynamical models can
be used to calibrate the \mlp\ for use with 1D models
\citep{LFS99}. Further steps in this direction were made by \citet{LAH02} who
demonstrated that this approach works well also with the 3D hydrodynamical
model of M-type dwarf.

In order to check the feasibility of such approach with the red giant model
studied here, we utilized the set of \LHD\ models calculated in the previous
section using different values of \mlp\ and plotted their entropy profiles
versus the optical depth in Fig.~\ref{f:stau}. Also shown there are the
temporally and spatially averaged specific entropy profiles of the
\xtmean{\mbox{3D}} model. Clearly, the 3D entropy profile shows hints of
an asymptotic flattening, similarly to what is seen in the 1D models. We
emphasize that horizontally resolved entropy profiles show a plateau at a
value indicated by $s_\mathrm{env}$ in Fig.~\ref{f:stau}.  Interpolation
between the asymptotic entropy values found in the 1D models with different
values of the mixing-length parameter and zero turbulent pressure yields
$\mlp\sim1.8$.

%The calibration of the mixing-length parameter in red giants could therefore
%be of interest and, ideally, should be done in a wider space of atmospheric
%parameters typical to these stars. While such endeavor would be clearly
%beyond the scope of this paper, it would be anyway interesting to see whether
%the approach proposed by \citet{LFS99} could be feasible with the 3D
%hydrodynamical model studied in this work.

One remark of caution is here again related to the turbulent
pressure. Evidently, one may get significantly different \mlp\ values for the
1D models characterized by different turbulent pressure, $P_{\rm turb}$
(Fig.~\ref{f:stau}). As we have seen in the previous section, there is no
straightforward procedure to calibrate turbulent pressure with the aid of
\xtmean{\mbox{3D}} models. Moreover, the net effect of the individual
spatially resolved entropy profiles will be always different from that given
by the \xtmean{\mbox{3D}} model. These facts once again stress the importance
and difficulties of the proper incorporation of turbulent pressure into the
atmospheric and evolutionary models of red giants.

\subsubsection{How realistic is our 3D hydrodynamical model?}

Obviously, the realism of the 3D hydrodynamical model discussed in this study
is limited by a number of the assumptions used in the modeling procedure. Some
issues need to be mentioned here.

Our model of a red giant has limited grid resolution of $150\times150$
horizontal grid points spread over $\sim1/6\times1/6$ of the stellar
surface. While such coverage and resolution may be sufficient to pin down the
general properties of atmospheric stratifications, substantially better
resolution would be needed to investigate the properties of atmospheric
motions on smaller scales. Moreover, the fraction of the surface area
covered by the model is already large and so the effects of sphericity
may become important. These issues should be addressed in future work.

As was described in Sect.~\ref{s:cobold_model}, we used monochromatic
opacities grouped into 5 opacity bins to reduce the computational work load.
Previous studies and tests that we did in the course of the present study have
demonstrated that opacity grouping yields a surprisingly good agreement with
the exact radiative transfer calculations obtained using opacity distribution
functions, ODFs \citep[see, e.g.,][]{N82,CAT07}. One has to be cautious,
however, since the effects related to molecular opacities may become
increasingly important in the outer atmospheres of red giants, and the opacity
grouping into small number of bins may be insufficient.

One more issue of concern may be related with the treatment of scattering. It
has been shown by \citet{CAT07,C08} that scattering treated as true absorption
may produce significantly warmer temperature profiles in the outer atmospheres
of red giants at low metallicities, compared to those that are obtained when
scattering is treated properly. The treatment of coherent and isotropic
scattering in the solution of radiative transfer equation was recently
implemented in the 3D hydrodynamical code {\tt BIFROST} by
\citet{HAC10}. While it seems that in general the continuum scattering does
not have large effect on the thermal structure in the atmospheric model of the
Sun, proper inclusion of scattering in the line blanketing may reduce the
temperature by $\sim350$\,K in the optically thin layers below $\log \tau_{\rm
  5000}\lesssim-4$ \citep{HAC10}. Obviously, this issue has to be addressed
properly in the future 3D hydrodynamical modeling of red giants, however, is
mainly an issue for metal-poor objects.

%------------------------------------------------------------------------
\section{Conclusions and outlook}
%------------------------------------------------------------------------

In this study we have utilized a 3D hydrodynamical model of a red giant
calculated with the \COBOLD\ code to investigate the influence of convection
on the properties of its atmospheric structures. The model had solar
metallicity and its atmospheric parameters, $T_{\rm eff}=3660$\,K and $\log
g=1.0$, were typical to those of a red giant located close to the RGB tip. We
also used a number of 1D atmosphere models calculated with the \LHD\ code
to compare the predictions of the 3D hydrodynamical model with those of
classical 1D models. The 1D \LHD\ model atmospheres shared the same opacities
and equation-of-state as used in the calculations of the 3D hydrodynamical
models and covered the same range in the optical depth. Thus, the comparison
was done in a strictly differential way, to ensure that the differences
revealed would reflect solely the differences related to dimensionality
between the two types of models.

Similarly to what is seen in dwarfs, the giant model predicts the existence of
pronounced surface granulation pattern with a white light intensity contrast
larger than in solar models ($\sim18\%$ versus $\sim15\%$, respectively). The
size of the granules relative to the available stellar surface is
significantly larger in the giant than in the Sun, with a typical ratio of
$\sim400$ and $\sim2\times10^6$ granules over surface area, respectively. This,
together with the larger intensity (and temperature) contrast in the giant
model suggests that convection may have substantially larger influence on the
observable properties (such as spectral energy distribution, photometric
colors) of red giants than in the Sun. Exploratory work in this direction
seems to confirm this prediction \citep{KLC09,KDI10}.

It is important to note that the atmosphere model displays a significant
activity of shock waves, especially in the outer atmosphere. In comparison
with the mild shocks in models of the Sun, with vertical and horizontal
velocities of up to Mach $\sim1.5$ and $\sim1.8$, respectively, the shocks in
the atmosphere of the giant are considerably stronger, with corresponding Mach
numbers of $\sim2.5$ and $\sim6$. We also find that the mean temperature of
the hydrodynamical model is slightly higher than that of the 1D model with
identical atmospheric parameters.

Turbulent pressure is considerably more important in the giant model than it
is in the Sun and may amount to $\sim35\%$ of the total pressure (i.e., gas
plus turbulent). This leads to a lifting of the outer atmospheric layers to
larger radii, besides, it would alter the pressure-temperature
relation. Interestingly, no combination of the mixing-length parameter and
turbulent pressure in the 1D models satisfactorily reproduce the mean
pressure-temperature relation of the hydrodynamical model.

We used the 3D hydrodynamical model to obtain the effective mixing-length
parameter for the classical 1D stellar atmosphere work. Interpolation between
the entropy profiles of the 1D models calculated with different mixing-length
parameters and zero turbulent pressure yields $\mlp=1.8$. This solution,
however, is not unique since identical mixing-length parameter may be obtained
for models characterized by different amount of turbulent pressure. The
question of how to treat the turbulent pressure in the 1D models of stellar
atmospheres and stellar evolution, therefore, still remains open. 
\change{In view of Fig.~\ref{f:tp} it appears unlikely that this can
  be addressed within the framework of MLT when assuming a sharp boundary of the
  convectively unstable region. In one way or another some ``softening'' by
  overshooting needs to be introduced. 3D models can give some quantitative
  guidance here. The indirect feedback of horizontal inhomogeneities on the vertical
  structure is also included in this way.}

Obviously, the 3D hydrodynamical stellar atmosphere model studied in this work
is rather an exploratory one. There is still some way to go for improving the
3D models so that they could match the realism of the current 1D stationary
stellar model atmospheres in the treatment of opacities, equation-of-state,
radiative transfer, and scattering. We believe, nevertheless, that the
advantages of the 3D hydrodynamical models over their classical 1D
counterparts are obvious, which warrants further efforts towards their firmer
implementation in the field of stellar atmosphere studies.

%------------------------------------------------------------------------
\acknowledgements
%------------------------------------------------------------------------

We warmly thank Violeta Ku\v{c}inskien\.{e}, Agn\.{e} Ku\v{c}inskait\.{e}, and
Simas Ku\v{c}inskas for their hospitality, repeatedly hosting the first author
of this article. We further thank Matthias Steffen for his help concerning the
evaluation of the radiative time scale in the non-grey case. HGL acknowledges
financial support from EU contract MEXT-CT-2004-014265 (CIFIST), and by the
Sonderforschungsbereich SFB\,881 ``The Milky Way System'' (subproject A4) of
the German Research Foundation (DFG). This work was in part supported by
grants of the Lithuanian Research Council (TAP-52 and MIP-101).

%------------------------------------------------------------------------
\bibliographystyle{aa}

%------------------------------------------------------------------------
\appendix
%------------------------------------------------------------------------

\section{Computation of the characteristic time scales\label{appendixa}}

The time scales in the red giant explored in this work (see Sect.~\ref{s:timescales}) were calculated following the prescriptions given in \citet{LAH02,LAH06}.

The radiative relaxation time, $t_\mathrm{rad}$, can be evaluated using the MLT formula

\beq
   t_\mathrm{rad} = \frac{\rho\cp\lmix\taueddy}{f_3\sigma T^3} \,
   \left( 1+\frac{f_4}{\taueddy^2}\right),
\label{e:tauradi}
\eeq

\noindent where $\rho$ is gas density, \cp\ the specific heat at constant pressure, $\lmix=\mlp\Hp$ the mixing-length, $\sigma$ Stefan-Boltzmann constant, $T$ gas temperature, $f_3=16$ and $f_4=2$ dimensionless constants in the MLT formulation of Mihalas \citep[see][]{LFS99}. The optical thickness of a convective element, $\tau_e$, is defined as $\taueddy\equiv\chi\rho\lmix$,
where $\chi$ is opacity.

It can be shown that in the case of non-grey radiative transfer with opacities grouped into $i$ opacity bins, Eq.~\eref{e:tauradi} can be re-written as

\beq
   t_\mathrm{rad} = \frac{\rho\cp\lmix}{f_3\sigma T^3} \,
   \left( \sum_{i} w_{\rm i}\frac{\tau_{\rm i}}{f_4 + \tau_{\rm i}^2}\right)^{-1},
\label{e:tauradbin}
\eeq

\noindent where the optical thickness of convective element is now different in each opacity bin

\beq
    \tau_{\rm i}\equiv\chi_{\rm i}\rho\lmix
\eeq

\noindent and  $\chi_{\rm i}$ is opacity in bin $i$. Eq.~\eref{e:tauradbin} was used to estimate the radiative time scale in the atmosphere of a red giant model studied here, with the weights $w_{\rm i}$ for the different opacity bins evaluated using

\beq
    w_{\rm i} = b_{\rm i} + \frac{T}{4}\frac{db_{\rm i}}{dT}
\eeq

\noindent where $b_{\rm i}=B_{\rm i}/B$ and $B=\frac{\sigma}{\pi}T^4$. The mixing-length parameter used in the calculations was set to \mlp=1.8, in accordance with the value derived in Sect.~\ref{s:mlp}.

%It was shown earlier by \citet{LAH02} that radiative time scales calculated using the Rosseland and Planck mean opacities may differ to two orders in magnitude in the outer atmospheres of M-type dwarfs. We therefore used both as Rosseland and Planck means to evaluate opacity in Equation~\ref{e:taueddy}, which therefore yielded two corresponding radiative time scales, $t_\mathrm{rad}^\mathrm{Ross}$ and $t_\mathrm{rad}^\mathrm{Planck}$.

The advection time scale, $t_\mathrm{adv}$, was estimated as time interval during which the convective element travels one mixing-length

\beq
   t_\mathrm{adv} = \frac{\lmix}{v_\mathrm{z}^\mathrm{rms}},
\eeq

\noindent where $v_\mathrm{z}^\mathrm{rms}$ is temporally and horizontally
averaged vertical RMS velocity.

The adiabatic Brunnt-V\"{a}is\"{a}l\"{a} period, $t_\mathrm{BV}$, was obtained using

\beq
   t_\mathrm{BV} = \frac{2\pi}{\sqrt{|\omega_\mathrm{BV}^2|}},
\eeq

\noindent with

\beq
   \omega_\mathrm{BV}^2 = \frac{\delta g}{\Hp}(\nabla_\mathrm{ad} - \nabla).
\eeq

\noindent where $\delta\equiv-(\frac{\partial\rho}{\partial T})_\mathrm{P}$ is the thermal expansion coefficient at constant pressure, $g$ gravitational acceleration, $\nabla_\mathrm{ad}$ and $\nabla$ the adiabatic and actual temperature gradients, respectively.

The Kelvin-Helmholtz time scale, $t_\mathrm{KH}$, was calculated using

\beq
   t_\mathrm{KH} = \frac{P \cp T}{g\sigma T_\mathrm{eff}^4},
\eeq

\noindent where $P$ is gas pressure.

Finally, the relative importance of radiative and advective heat transport was estimated using the P\'{e}clet number

\beq
   {\rm Pe}\equiv \frac{t_\mathrm{rad}}{t_\mathrm{adv}}.
\eeq

\section{Changes of the radiative equilibrium temperature due to spatial
  brightness fluctuations\label{appendixb}}

In Sect.~\ref{s:propaver} we found a slight ($\sim 20$\,K) increase of the mean
atmospheric temperature of the 3D model in the upper atmosphere relative to
the radiative equilibrium temperature of a plane-parallel 1D model
atmosphere. 3D and 1D model share the same effective temperature (and
gravity). We want to demonstrate here that this temperature increase in the 3D
model can be interpreted as a change of the radiative equilibrium
temperature caused by the spatial fluctuations of the brightness in deeper
atmospheric layers due to the convection pattern present in the 3D model. To
this end, we develop a simple model describing the establishment of the radiative
equilibrium temperature when brightness fluctuations are present.

The radiative equilibrium temperature is established by the balance of
absorbed radiation coming from deeper layers and the radiation emitted according to
the local temperature. We consider the radiation field as simply given by the
Kirchhoff-Planck function $\Blam(T)$ of radiation temperature $T$. In the
plane-parallel case the balance between emission at local temperature $T_1$
and absorption of radiation coming from the deeper atmospheric layers with
radiation temperature $T_0$ can be written as
\beq
\intlam \klam(T_1)\Blam(T_1) = \fom\intlam \klam(T_1)\Blam(T_0)
\label{e:tradeq}
\eeq
where \klam\ is the opacity and \fom\ the solid angle subtended by the
radiating deeper layers. We envision this to be the ``stellar surface'', and
$T_0$ is close to the effective temperature. For brevity, we did not
explicitely note the pressure dependence of the opacity $\klam(P_1,T_1)$ in
the above equation. All wavelength integrals in this section should be taken
from zero to infinity. \fom\ is $\frac{1}{2}$ in the case of no
limb-darkening, decreasing to smaller values as $\fom =
\frac{1}{2}(1-\frac{1}{2} a)$ for a limb-darkening being described by a linear
limb-darkening law with coefficient~$a$. We now introduce brightness
fluctuations by considering a two component model for the granulation having
equal surface area fractions and radiation temperatures $T_0+\dT_0$ and
$T_0-\dT_0$. The total emitted flux should correspond to the unperturbed
situation so that we have to fulfill the normalization condition
\beq
\fb \intlam \frac{1}{2}\left\{\Blam(T_0+\dT_0)+ \Blam(T_0-\dT_0)\right\}=
\intlam \Blam(T_0),
\label{e:fb1}
\eeq
introducing the normalization factor \fb. The temperature fluctuations
lead to a ``hardening'' of the radiation field, i.e., a
shift of the flux towards the blue part of the spectrum. Since
$\intlam\Blam(T)=\frac{\sigma}{\pi} T^4$ ($\sigma$ indicates
Stefan-Boltzmann's constant), Eq.~\eref{e:fb1} could be solved exactly for
$\dT_0$. However, here we are content with an approximate -- and much simpler
-- solution for the case $\dT_0 \ll T_0$ and obtain to leading order
\beq
\fb \approx \frac{1}{1+6\left(\frac{\dT_0}{T_0}\right)^2} .
\eeq

\begin{figure}
\centering
\includegraphics[width=\columnwidth]{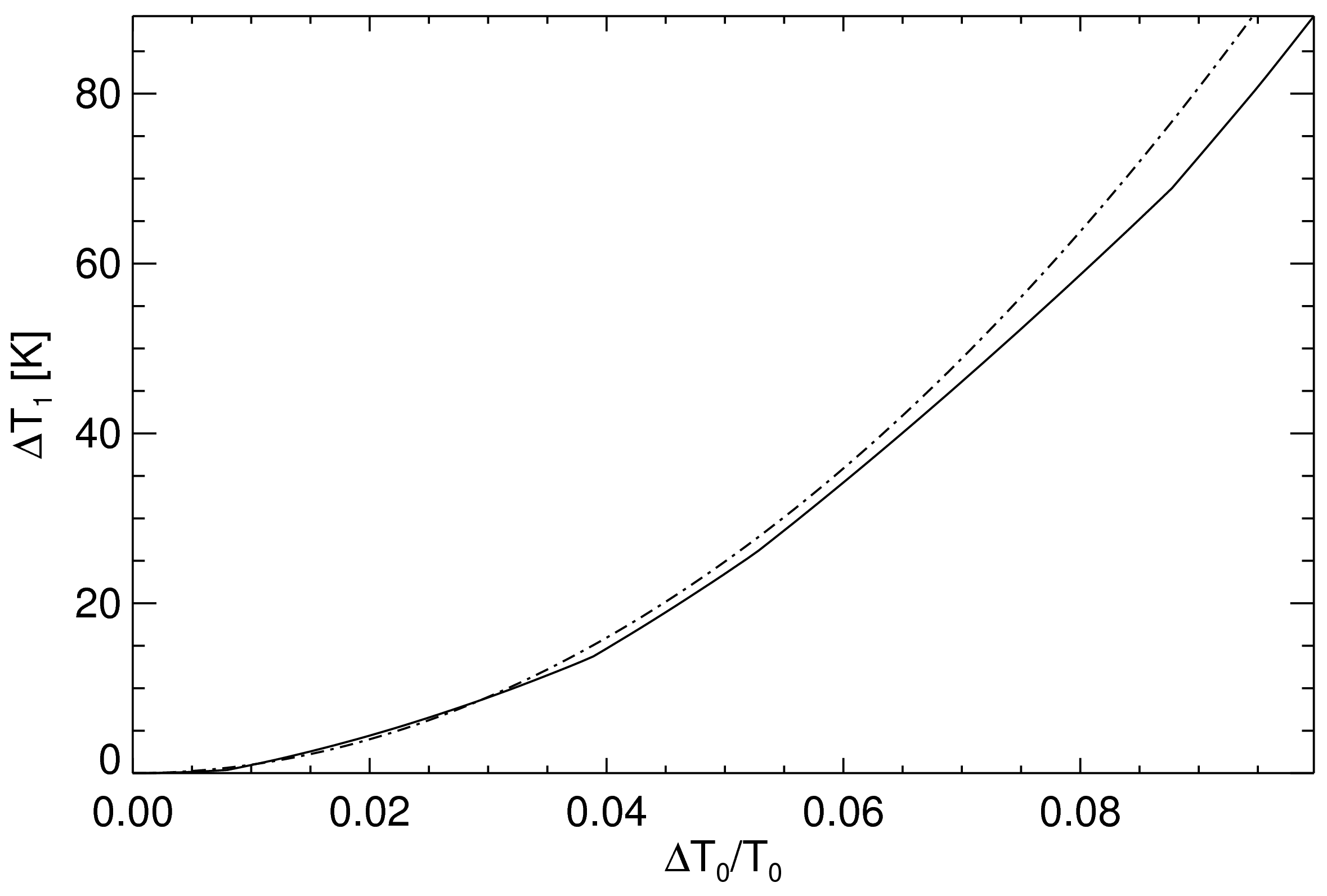}
\caption{Solid line: change of the radiative equilibrium temperature $\dT_1$ due to
  horizontal temperature fluctuations according Eqs.~\eref{e:fb1}
  and~\eref{e:tradeq2}. Dash-dotted line: dependence as given by the
  linearized solution~\eref{e:dT1}. Further details see text.}
\label{f:dt1}
\end{figure}

We now evaluate the change in the radiative equilibrium temperature~$\dT_1$
relative to the plane-parallel situation due to brightness fluctuations
associated with temperature changes of size~$\dT_0$. We use the
ansatz~\eref{e:tradeq} to obtain for the relation between emission and
absorption at a location in the optically thin layers of the atmosphere
\beq
\intlam \klam\Blam(T_1+\dT_1) = \fb\fom\intlam \klam
\frac{1}{2}\left\{\Blam(T_0+\dT_0)+ \Blam(T_0-\dT_0)\right\} .
\label{e:tradeq2}
\eeq
Expanding for $\dT_1 \ll T_1$ results to leading order in
\beq
\dT_1\approx 6 \fom\left(\frac{\dT_0}{T_0}\right)^2
\frac{\intlam\,\klam(T_1)\left[\frac{1}{12}\left.\frac{d^2\Blam}{dT^2}\right|_{T_0}T_0^2-\Blam(T_0)\right]}{\intlam\,\klam(T_1)\left.\frac{d\Blam}{dT}\right|_{T_1}} .
\label{e:dT1}
\eeq
Equation~\eref{e:dT1} shows that the influence of horizontal inhomogeneities
on the radiative equilibrium temperature is small since $\dT_1$ goes as the
square of the temperature fluctuations. Moreover, $\dT_1$ is controlled by the
complex dependence of the opacity~$\klam$ on wavelength: $\dT_1$ vanishes if
the opacity is wavelength-independent. To account for the wavelength
dependence of the opacity the integrals in Eq.~\eref{e:dT1} have to be
evaluated numerically. In doing so, we decided to go one step back and
evaluate the fundamental equations~\eref{e:fb1} and~\eref{e:tradeq2} before
expanding for small $\dT_0$ and $\dT_1$. Moreover, we used the binned
opacities as applied in the radiative transfer of the 3D and 1D model to
approximate the integrals improving consistency of our approximate analytical
model with the detailed models. A solution for $\dT_1$ for given $\dT_0$ can
be found iteratively. For completeness, we also calculated the dependence
according Eq.~\eref{e:dT1}, again, evaluating the integrals using the binned
opacities.

Figure~\ref{f:dt1} depicts the result for $T_0=3660\,\mbox{K}$,
$T_1=2900\,\mbox{K}$, and $P_1=100\,\mbox{dyn cm}^{-2}$. $T_0$ corresponds to
the effective temperature of the detailed numerical models, $T_1$ and $P_1$ to
the conditions prevailing around optical depth $\log \tau_{\rm Ross} \sim
-3$. $T_1$ was adjusted by setting the solid angle fraction to $\fom=0.28$
(indicating a rather strong limb-darkening which is actually present).  The
model indeed predicts a heating of the outer atmosphere while a cooling would
be also possible depending on the wavelength-dependence of the opacity.  A
white light intensity contrast of about 18\,\% as suggested by
Fig.~\ref{f:intensmaps} corresponds to a temperature fluctuation of
4.5\,\%. At this level of fluctuations our simple analytical model predicts an
increase of the radiative equilibrium temperature in the optically thin layers
close to the $\sim$20\,K found in the 3D model (see Fig.~\ref{f:dt}). Hence,
we consider it plausible that the the temperature increase in the optically
thin layers in the 3D model relative to 1D plane-parallel models is a
consequence of horizontal brightness fluctuations associated with granulation.

\end{document}